\documentclass[12pt,a4paper]{article}
\usepackage[english]{babel}
\usepackage[numbers,comma,square,compress]{natbib}
\usepackage{amsmath,amssymb,amsfonts}
\usepackage{bm,enumerate}
\usepackage{ifpdf}
\ifpdf  
  \usepackage[pdftex,a4paper,margin=24mm]{geometry}
  \usepackage[pdftex]{hyperref}
\else  
  \usepackage[dvips,a4paper,margin=24mm]{geometry}
  \usepackage[dvips]{hyperref}
\fi

\numberwithin{equation}{section}

\newcommand{\email}[1]{\href{mailto:#1}{#1}}
\newcommand{\EH}{Einstein-Hilbert }
\newcommand{\ADM}{Arnowitt-Deser-Misner }
\newcommand{\HL}{Ho\v{r}ava-Lifshitz }
\newcommand{\GBC}{Gauss-Bonnet-Chern }
\newcommand{\SWeyl}{S_\mathrm{Weyl}}
\newcommand{\SWeylLambda}{S_\mathrm{Weyl,\Lambda}}
\newcommand{\Ssurface}{S_\mathrm{surf}}
\newcommand{\Hsurface}{H_\mathrm{surf}}
\newcommand{\nn}{\nonumber}

\newcommand{\p}{\partial}
\newcommand{\pb}[1]{\left\{#1\right\}}
\newcommand{\db}[1]{\left\{#1\right\}_{\mathrm{D}}}
\newcommand{\cM}{\mathcal{M}}
\newcommand{\cH}{\mathcal{H}}
\newcommand{\mH}{\mathcal{H}}
\newcommand{\cL}{\mathcal{L}}

\newcommand{\cK}{\mathcal{K}}
\newcommand{\cB}{\mathcal{B}}

\newcommand{\cG}{\mathcal{G}}
\newcommand{\bcG}{\bar{\mathcal{G}}}

\newcommand{\cN}{\mathcal{N}}
\newcommand{\cP}{\mathcal{P}}
\newcommand{\bcP}{\bar{\mathcal{P}}}

\newcommand{\cQ}{\mathcal{Q}}

\newcommand{\bx}{x}
\newcommand{\by}{y}
\newcommand{\bz}{z}
\newcommand{\bn}{\bm{n}}

\newcommand{\projector}[2]{h^{#1}_{\phantom{#1}#2}}
\newcommand{\sR}[1][3]{{}^{(#1)}\!R}
\newcommand{\sC}[1][3]{{}^{(#1)}\!C}

\newcommand{\projected}[1]{{}_\perp\!{#1}}
\newcommand{\pR}{\projected{R}}
\newcommand{\pC}{\projected{C}}
\newcommand{\Kt}[1][2]{{}^{(#1)}\!K}

\newcommand{\Lie}[1]{\mathcal{L}_{#1}}
\newcommand{\MC}[1]{\Phi\bigl[#1\bigr]}
\newcommand{\HC}[1]{\cH_0[#1]}

\begin{document}
\begin{center}
{\Large Hamiltonian analysis of curvature-squared gravity\\
\vspace{.4em}
with or without conformal invariance}

\vspace{1em}
Josef Kluso\v{n}$\,^a$,
Markku Oksanen$\,^b$,
Anca Tureanu$\,^b$\,\footnote{Email addresses:
\email{klu@physics.muni.cz} (J. Kluso\v{n}),
\email{markku.oksanen@helsinki.fi} (M. Oksanen),
\email{anca.tureanu@helsinki.fi} (A. Tureanu)}\\
\vspace{1em}
$^a$\textit{Department of Theoretical Physics and Astrophysics, Faculty
of Science,\\
Masaryk University, Kotl\'a\v{r}sk\'a 2, 611 37, Brno, Czech Republic}\\
\vspace{.3em}
$^b$\textit{Department of Physics, University of Helsinki, P.O.
Box
64,\\ FI-00014 Helsinki, Finland}\\
\end{center}

\begin{abstract}
We analyze gravitational theories with quadratic curvature terms,
including the case of conformally invariant Weyl gravity, motivated by
the intention to find a renormalizable theory of gravity in the
ultraviolet region, yet yielding general relativity at long distances.
In the Hamiltonian formulation of Weyl gravity, the number of local
constraints is equal to the number of unstable directions in phase
space, which in principle could be sufficient for eliminating the
unstable degrees of freedom in the full nonlinear theory. All the other
theories of quadratic type are unstable -- a problem appearing as ghost
modes in the linearized theory.

We find that the full projection of the Weyl tensor onto a
three-dimensional hypersurface contains an additional fully traceless
component, given by a quadratic extrinsic curvature tensor. A certain
inconsistency in the literature is found and resolved: when the
conformal invariance of Weyl gravity is broken by a cosmological
constant term, the theory becomes pathological, since a constraint
required by the Hamiltonian analysis imposes the determinant of the
metric of spacetime to be zero. In order to resolve this problem by
restoring the conformal invariance, we introduce a new scalar field that
couples to the curvature of spacetime, reminiscent of the introduction
of vector fields for ensuring the gauge invariance.
\end{abstract}


\section{Introduction}
One of the most interesting theories of gravity is the Weyl gravity
\cite{Weyl:1918}, whose action is defined by the square of the Weyl
tensor, $S=-\frac{1}{4}\int d^4x\sqrt{-g}C_{\mu\nu\rho\sigma}
C^{\mu\nu\rho\sigma}$.
The intriguing property of this theory is its invariance under the local
conformal transformation of the metric,
$g_{\mu\nu}\rightarrow\Omega^2(x)g_{\mu\nu}$, making it consequently
insensitive to the angles. Furthermore, it is a power-counting
renormalizable theory of gravity thanks to the presence of higher-order
derivatives in the Lagrangian.
Hence it can be considered as an ultraviolet completion of gravity.
For a review, see \cite{Mannheim:2012} and references therein. More
generally, perturbative renormalization of general relativity requires
us to add to the \EH Lagrangian invariant counterterms that are
quadratic in curvature \cite{tHooft:1974,Deser:1974,Deser:1974b}.
Furthermore, Weyl gravity is also useful for the supergravity
construction \cite{Bergshoeff:1980is,deWit:1980tn} and it also emerges
from the twistor string theory \cite{Berkovits:2004jj}.

Weyl gravity is a special case of higher-derivative theories of gravity
which have been extensively studied, especially in the case
of three dimensions. One such example is the new massive gravity in
three dimensions \cite{Bergshoeff:2009hq}, whose Lagrangian includes
quadratic curvature terms that contain four-derivative contributions.
There also exists a recent example of four-dimensional higher-derivative
gravity which is a combination of the Einstein gravity with cosmological
constant, together with the contribution of the Weyl tensor squared term.
An appropriate fine-tuning of the cosmological constant and the
coefficient of the Weyl term produces the so-called critical gravity,
where the additional massive spin-2 excitations around an anti-de Sitter
background become massless \cite{Lu:2011zk}.
The relation of such critical gravity to conformal gravity in
four-dimensional spacetime has been studied in \cite{Lu:2011ks}.
Recently it has been proposed that one can obtain solutions of
four-dimensional Einstein gravity with cosmological constant by
introducing a simple Neumann boundary condition into the conformally
invariant Weyl gravity \cite{Maldacena:2011}. In a somewhat similar
fashion, it has been argued that one can obtain ghost-free
four-dimensional massive gravity by introducing Dirichlet boundary
conditions into curvature-squared gravity on an asymptotically de Sitter
spacetime \cite{Park:2012}. Weyl action is also an important object
in recent proposals, that relates the conformal symmetry group,
gravity and particle physics \cite{Hooft:2010ac,tHooft:2011aa}.
It is possible that this idea is closely related to the earlier proposal
by Sakharov on the generation of both \EH action and higher-order
curvature terms from the quantum fluctuations of vacuum when space is
curved \cite{Sakharov:1968}. For further works on the generation of
Einstein gravity as a quantum effect, see
\cite{Adler:1982,Zee:1983,Smilga:1984}.

The inclusion of higher-order curvature terms is generally motivated by
string theory, since such terms are known to appear in the low energy
limit \cite{Boulware:1985wk,Fradkin:1985ys}.

On the other hand, the price that we have to pay in order to achieve a
renormalizable theory of gravity is the inclusion of extra
gravitational degrees of freedom. They appear because of the
higher-order time derivatives present in the Lagrangian. Moreover,
those extra degrees of freedom often have negative kinetic terms, and
are usually referred to as ghosts when the linearized theory is
quantized. Typically, theories with ghosts are considered to be
inconsistent, since they are either violently unstable or nonunitary,
depending on whether the states associated with the higher-derivative
degrees of freedom are considered to possess negative energy or positive
energy but indefinite norm. However, there are attempts how to
resolve this problem as for example in \cite{Bender:2007wu}; see also
\cite{Smilga:2008pr,Smilga:1995}.

It is important to stress that there is an alternative way to construct
a renormalizable theory of gravity, which is called \HL gravity
\cite{Horava:2008ih,Horava:2009uw}. It achieves renormalizability via
reduction of the gauge symmetry. Since the theory is invariant
under foliation preserving diffeomorphisms, one can exclude higher-order
time derivatives from the action, avoiding the ghost problem. The lack
of full diffeomorphism invariance has significant consequences for the
structure of the theory. For reviews, see
\cite{Sotiriou:2010wn,Visser:2011mf}. The Hamiltonian formulation of \HL
gravity has been studied particularly in
\cite{Henneaux:2010,Donnelly:2011}. For a review, see
\cite{Oksanen:thesis}.
Generalized \HL gravitational theories were proposed and analyzed in
\cite{Kluson:2009xx,CNOOT:2010,CCNOOT:2010,COT:2010}.
Proposals of covariant alternatives to \HL gravity were presented in
\cite{Nojiri:2010tv,Kluson:2011rs} and their Hamiltonian structure was
studied in \cite{Chaichian:2011,Chaichian:2012md}.

Previous lines suggest that gravitational theories involving
higher-order curvature terms are very interesting and deserve to be
studied from different points of view. An important question is to
understand their Hamiltonian dynamics. The strong coupling limit of
conformal gravity was considered first in \cite{Kaku:1983}.
The Hamiltonian formulation of the higher-derivative theories of gravity
(up to quadratic curvature terms) was performed in
\cite{Boulware:1984,Buchbinder:1987}, and later also considered in
\cite{Demaret:1995,Querella:1998}. The
Hamiltonian analysis of more general $f(\mathrm{Riemann})$ theories of
gravity was considered in \cite{Deruelle:2009zk}.

The main goal of the present paper is to perform the Hamiltonian
analysis of the curvature-squared theories of gravity in greatest
details. The previous analyses have not been complete in all respects,
and due to the new interest in higher-derivative gravity they deserve a
new detailed treatment. Specifically, we are interested in the structure
of constraints which crucially depends on the values of the parameters
that appear in the action, what will be introduced in the next section.
It turns out that each case requires a separate treatment since the
nature of various constraints and the number of physical degrees
of freedom depend on the value of the parameters. The theory with
most symmetries is the Weyl gravity. A surprising situation occurs when
Weyl gravity is supplemented with a nonzero cosmological constant.
It obviously breaks conformal invariance, but diffeomorphism invariance
is retained. There is no evident reason why we could not include such a
constant term into the action. However, the investigation of the structure
of constraints reveals that the theory becomes inconsistent: the
requirement of the preservation of a secondary constraint leads to the
condition that the determinant of the metric of spacetime should be
equal to zero, which is satisfied when either the lapse $N$ or the
determinant of the three-dimensional metric $h$ is zero,
$\sqrt{-g}=N\sqrt{h}=0$. We analyze this situation further by
introducing a scalar field $\phi$ into the action in order to regain the
conformal invariance, following
\cite{Antoniadis:1984kd,Antoniadis:1984br}.
The scalar field has to couple to the scalar curvature and its kinetic
term must have the wrong sign in order to obtain the Weyl invariant
action. In the Hamiltonian analysis, we obtain a first-class
constraint that is the generator of the Weyl symmetry. This symmetry
can be gauge fixed by imposing the condition $\phi=\mathrm{const}$.
When we insert this condition into the action, we obtain the standard
Einstein-Hilbert term with the condition that the original kinetic term
for $\phi$ corresponds to the ghostlike degree of freedom.

\section{Gravitational action with quadratic curvature terms}
\label{sec2}
\subsection{Action and its physical degrees of freedom}
We consider the generally covariant theory of gravity in
four-dimensional spacetime whose action consists of the \EH part,
the cosmological constant and the quadratic curvature terms.
The gravitational action reads
\begin{equation}\label{S_C}
S_C = \int d^4x\sqrt{-g}\left[ \Lambda + \frac{R}{2\kappa}
- \frac{\alpha}{4}C_{\mu\nu\rho\sigma}C^{\mu\nu\rho\sigma}
+ \frac{\beta}{8}R^2 + \gamma G \right],
\end{equation}
where $\kappa$, $\alpha$, $\beta$ and $\gamma$ are coupling constants.
We consider all four-dimensional integrals to be taken over the whole
spacetime $\cM$.
The Weyl tensor is defined as
\begin{equation}\label{Weyl-tensor}
C_{\mu\nu\rho\sigma} = R_{\mu\nu\rho\sigma} - \frac{2}{(d-2)}\left(
g_{\mu[\rho}R_{\sigma]\nu} - g_{\nu[\rho}R_{\sigma]\mu} \right)
+ \frac{2}{(d-1)(d-2)} g_{\mu[\rho}g_{\sigma]\nu}R \,,
\end{equation}
where $d$ is the dimension of spacetime. The Weyl tensor is by definition
the traceless part of the Riemann tensor.
In the last term of the action \eqref{S_C}, $G$ is the \GBC curvature term
\begin{equation}\label{GBC}
G=R_{\alpha\beta\gamma\delta}R_{\mu\nu\rho\sigma}
\epsilon^{\alpha\beta\mu\nu}\epsilon^{\gamma\delta\rho\sigma}
=R_{\mu\nu\rho\sigma}R^{\mu\nu\rho\sigma}
- 4R_{\mu\nu}R^{\mu\nu} + R^2 \,.
\end{equation}
In four-dimensional spacetime, its integral $\int d^4x\sqrt{-g}G$
becomes a topological invariant which is proportional to the Euler
characteristic of the spacetime manifold. Since we consider smooth
variations of spacetime, which do not change its topology, the \GBC part
of the action can be regarded as a constant. Hence we can drop it.

The Weyl tensor squared can be written as
\begin{equation}\label{Weyl.squared}
C_{\mu\nu\rho\sigma}C^{\mu\nu\rho\sigma} = 2\left(
R_{\mu\nu}R^{\mu\nu} - \frac{1}{3}R^2 \right) + G \,.
\end{equation}
Thus the action \eqref{S_C} becomes
\begin{equation}\label{S_R}
S_R = \int d^4x\sqrt{-g}\left[ \Lambda + \frac{R}{2\kappa}
- \frac{\alpha}{2} R_{\mu\nu}R^{\mu\nu} + \left( \frac{\alpha}{6}
+\frac{\beta}{8} \right) R^2 + \left( \gamma - \frac{\alpha}{4} \right)G
\right].
\end{equation}
The actions \eqref{S_C} and \eqref{S_R} are of course identical.
However, when we discard the \GBC topological invariant term
in both actions, $\gamma\int d^4x\sqrt{-g}G$ and $(\gamma-\alpha/4)
\linebreak\int d^4x\sqrt{-g}G$, respectively, the resulting actions
differ by a multiple of the said invariant, namely by $-(\alpha/4)\int
d^4x\sqrt{-g}G$ . Even though this topological invariant term has no
impact on the physical dynamics of the theory, it affects the
Hamiltonian formulation of the theory in a significant way. Namely, the
structure of the constraints of the theory is considerably simpler for
the action \eqref{S_C} than for the action \eqref{S_R}, when both
actions are considered without the explicit invariant term $\int
d^4x\sqrt{-g}G$. In other words, the topological invariant part of the
Weyl tensor squared \eqref{Weyl.squared} action simplifies the
Hamiltonian analysis significantly. 

The quadratic curvature terms are known to render the theory
renormalizable when the cosmological constant is absent
\cite{Stelle:1977}. The theory is also known to possess the property of
asymptotic freedom \cite{Fradkin:1982}. For nonvanishing couplings, the
action \eqref{S_C} contains eight local degrees of freedom
\cite{Stelle:1977,Stelle:1978}. On the Minkowski background, two
degrees of freedom are associated with the usual massless spin-2
graviton, five modes are associated with a massive spin-2 excitation,
and one with a massive scalar. Moreover, the massive spin-2 component
carries negative energy, which implies that the theory is unstable.
Alternatively, the negative energy states can be regarded as positive
energy states with indefinite norm, what leads to the violation of
unitarity.

Although pure curvature-squared gravity without the \EH part admits
standard vacuum solutions, asymptotically flat solutions do not couple
to a positive definite matter distribution
\cite{Deser:1974b,Stelle:1978}, e.g., in the case of the Schwarzshild
solution. For that reason the \EH part of the action is necessary in the
infrared region. As we already noted above, the \EH action can be
generated by quantum effects whenever the spacetime is allowed to be
curved \cite{Sakharov:1968,Adler:1982,Zee:1983,Smilga:1984}.

Some of the extra degrees of freedom can be removed by certain choices
of the coupling constants.
For $\Lambda=0$, $\kappa^{-1}=\beta=\gamma=0$ and $\alpha=1$, the
action \eqref{S_C} becomes the conformally invariant action of Weyl
gravity
\begin{equation}\label{S.Weyl}
\SWeyl =-\frac{1}{4}\int d^4x\sqrt{-g}
C_{\mu\nu\rho\sigma}C^{\mu\nu\rho\sigma}\,.
\end{equation}
The additional local symmetry under conformal transformations removes
one degree of freedom, but it is not sufficient for removing all the
ghosts, namely the negative energy spin-2 excitations.
On the Minkowski background, the six degrees of freedom are associated
with ordinary massless spin-2 and spin-1 excitations, and with a
massless spin-2 ghost \cite{Riegert:1984}.

On the cosmologically relevant anti-de Sitter backgrounds, the extra
gravitational degrees of freedom can appear as a partially massless
spin-2 field \cite{Maldacena:2011,Deser:2012qg}. This is because the
conventional connection between gauge invariance, masslessness and
propagation on null cones holds generically only in flat
four-dimensional spacetime \cite{Deser:1983mm}. On (anti-)de Sitter
backgrounds, higher spin fields ($s>1$) can become partially massless,
carrying a number of degrees of freedom that is between the extremes of
flat space, $2s+1$ for massive and $2$ for massless fields. This
requires that the mass is appropriately tuned with respect to the
cosmological constant \cite{Deser:2001pe}.

On the other hand, setting $\alpha=0$ in the action removes the
negative energy spin-2 component, leaving only the extra scalar degree
of freedom. The result would indeed be the simplest case of $f(R)$
gravity. But then renormalizability is lost. We consider the potentially
renormalizable theories exclusively in this paper.
Hence we assume $\alpha\neq0$.

\subsection{Equations of motion}
The variation of the gravitational action with respect to the metric of
spacetime $g^{\mu\nu}$ leads to the following equations of motion
\begin{multline}\label{EoM}
 -g_{\mu\nu}\Lambda + \frac{1}{\kappa}\left(
R_{\mu\nu}-\frac{1}{2}g_{\mu\nu}R \right)
+\alpha\left( 2\nabla^\rho\nabla^\sigma C_{\rho\mu\nu\sigma}
+C_{\rho\mu\nu\sigma}R^{\rho\sigma} \right)\\
+ \frac{\beta}{2}\left[ R\left( R_{\mu\nu}-\frac{1}{4}g_{\mu\nu}R
\right) - \nabla_\mu\nabla_\nu R + g_{\mu\nu}\nabla^\rho\nabla_\rho R
\right] = T_{\mu\nu} \,,
\end{multline}
where the energy-momentum tensor of matter is defined as
\begin{equation}
T_{\mu\nu}=-\frac{2}{\sqrt{-g}}\frac{\delta
S_\mathrm{matter}}{\delta g^{\mu\nu}}\,.
\end{equation}
Matter is assumed to be coupled minimally to the metric of spacetime.
The trace of the equations of motion is a linear inhomogeneous
second-order differential equation for the scalar curvature,
\begin{equation}
-4\Lambda - \frac{1}{\kappa}R + \frac{3\beta}{2}\nabla^\mu\nabla_\mu R
= T\,,
\end{equation}
with no contribution from the Weyl gravity part of the action.
The boundary terms which arise from the variation of the action are
discussed next.

\subsection{Boundary surface terms}\label{sec2.3}
We require that the solutions of Einstein field equations be
extrema of the \EH action when only the variation of the metric (and not
its derivatives) is fixed to zero on the boundary of spacetime.
In general relativity, we cannot fix derivatives of the variation
of the metric on the boundary, because that would overconstrain the
system. Therefore, we have to discuss the surface terms that arise in
the variation of the action with respect to the metric of spacetime.

The variations of the metric are required to vanish on the boundary of
spacetime, $\delta g_{\mu\nu}=0$. In fact, it is sufficient to fix only
the variation of the induced metric $\gamma_{\mu\nu}$ on the boundary of
spacetime, $\delta\gamma_{\mu\nu}=0$, while leaving the variation
normal to the boundary free. This is because a variation of the action
is invariant under diffeomorphism gauge transformations and there
always exists a gauge transformation $\nabla_{(\mu}\xi_{\nu)}$ with
$\xi_{\nu}=0$ on the boundary, which transforms
$\delta g_{\mu\nu}$ into $\delta\gamma_{\mu\nu}$.

Whenever the integrand of a surface term is proportional to $\delta
g_{\mu\nu}$ or $\delta\gamma_{\mu\nu}$ on the boundary of spacetime, the
surface term vanishes. Thus we only need to consider surface terms that
involve derivatives of the variation of the metric.

The variation of the Einstein-Hilbert action with respect to the metric
includes a nonvanishing surface term on the boundary of spacetime,
which can be written as
\begin{equation}\label{EH-surfaceterm}
-\frac{1}{2\kappa}\oint_{\p\cM}d^3x\sqrt{|\gamma|}\,r_\mu\left(
\nabla_\nu\delta g^{\mu\nu} -g_{\nu\rho}\nabla^\mu\delta g^{\nu\rho}
\right)
= -\frac{1}{\kappa}\oint_{\p\cM}d^3x\sqrt{|\gamma|}\delta K\,,
\end{equation}
where $\p\cM$ is the boundary of spacetime $\cM$, natural
integration measure on $\p\cM$ is assumed, $r^\mu$ is the
outward-pointing unit normal to the boundary (with norm
$\varepsilon=r_\mu r^\mu=\pm1$),
$\gamma_{\mu\nu}=g_{\mu\nu}-\varepsilon r_\mu r_\nu$ is the
induced metric on the boundary, and $\delta K$ is the variation of the
trace of the extrinsic curvature of the boundary of spacetime,
$K=\nabla_\mu r^\mu$.
In order to obtain a variational principle consistent with Einstein
equations when only the variation of the metric (and not its derivatives)
is fixed to zero on the boundary $\p\cM$, we add the surface term
$\frac{1}{\kappa}\oint_{\p\cM}d^3x\sqrt{|\gamma|}K$ into
the Einstein-Hilbert action, so that the surface term in the variation
of the original action gets canceled:
\begin{equation}\label{S.EH}
S_\mathrm{EH}=\frac{1}{2\kappa}\int d^4x\sqrt{-g}R
+\frac{1}{\kappa}\oint_{\p\cM}d^3x\sqrt{|\gamma|}K\,.
\end{equation}
This completion to \EH action was originally found in \cite{York:1972}
and later considered in \cite{Gibbons:1977}. It is regarded as the
complete standard action of general relativity.

As long as the geometry of spacetime is spatially compact, the action
\eqref{S.EH} is well defined. But in spatially noncompact spacetimes,
the action diverges. Then one must choose a reference background,
including the metric of spacetime $g_0$ and matter fields $\psi_0$, and
then define the physical action as the difference of the variable action
compared to the action of the fixed background \cite{Hawking:1995fd}:
\begin{equation}
S_{\mathrm{phys}}[g,\psi]=S[g,\psi]-S[g_0,\psi_0]\,.
\end{equation}
This physical action is finite if we require that the field variables
and the reference fields induce the same field configuration on the
boundary of spacetime (particularly at spatial infinity).

Does the variation of the curvature-squared part of the action
\eqref{S_C} yield extra surface terms? The variation of the action
indeed contains nonvanishing surface terms. The first one is obtained
from the $R^2$ part of the action in a similar way as in the case of \EH
action
\begin{equation}\label{Rsquared-surfaceterm}
-\frac{\beta}{4}\oint_{\p\cM}d^3x\sqrt{|\gamma|}\,Rr_\mu \left(
\nabla_\nu\delta g^{\mu\nu} -g_{\nu\rho}\nabla^\mu\delta g^{\nu\rho}
\right)
= -\frac{\beta}{2}\oint_{\p\cM}d^3x\sqrt{|\gamma|}R\delta K \,.
\end{equation}
The Weyl gravity part of the action implies the second surface term as
\begin{multline}\label{Weyl-surfaceterm}
\alpha\oint_{\p\cM}d^3x\sqrt{|\gamma|}\,r_\mu \left[
R_{\nu\rho}\nabla^\rho \delta g^{\mu\nu} -\frac{1}{2}\left(
R_{\nu\rho}-\frac{2}{3}g_{\nu\rho}R \right) \nabla^\mu\delta g^{\nu\rho}
\right.\\
-\left.\frac{1}{2}R^{\mu\nu}g_{\rho\sigma}\nabla_\nu\delta
g^{\rho\sigma} -\frac{1}{3}R\nabla_\nu\delta g^{\mu\nu}  \right].
\end{multline}
In general, it appears to be impossible to write either of these
boundary contributions as a variation of a functional on the boundary of
spacetime. Some cases of very high level of symmetry, e.g. maximally
symmetric spacetime, might be an exception.
Although in general relativity we cannot fix the covariant derivatives
of the variation of the metric on the boundary, we are now considering a
higher-derivative theory, where imposing boundary conditions on the
derivatives of the variation might be both permitted and natural,
because the metric carries extra degrees of freedom due to the
higher-order time derivatives.
We shall postpone the final discussion on surface terms until
Sec.~\ref{sec3.4}, where the theory is given in a first-order form
using \ADM formulation generalized for higher-derivative theory.
Surface terms arising in Hamiltonian formalism are discussed in
Sec.~\ref{sec4}.

The variation of the \GBC topological invariant vanishes identically when the
spacetime has no boundary. When the spacetime has boundary, the
variation contains a surface term that turns out to be a variation of a
functional on the boundary of spacetime. That boundary term can then be
added into the term $\int d^4\sqrt{-g}G$, thus obtaining a true
topological invariant whose variations vanish identically.
In the presence of boundaries, the topological invariant can be written
as
\begin{multline}\label{GBCwithBoundary}
\int d^4x\sqrt{-g}G + 4\oint_{\p\cM}d^3x\sqrt{|\gamma|}\left( RK
-2R_{\mu\nu}\gamma^{\mu\nu}K
+2R_{\mu\nu\rho\sigma}K^{\mu\rho}\gamma^{\nu\sigma} \right.\\
-\left.\frac{4}{3}K_{\mu\nu}K^\mu_{\phantom\mu\rho}K^{\nu\rho}
+K_{\mu\nu}K^{\mu\nu}K -\frac{2}{3}K^3 \right)
= -32\pi^2\chi(\cM) \,,
\end{multline}
where $K_{\mu\nu}$ denotes the extrinsic curvature on the boundary
of spacetime. The Euler characteristic of spacetime $\cM$ is denoted by
$\chi(\cM)$. For a brief review of \GBC theorem, see, e.g.,
\cite{Alty:1995}.

\section{First order Arnowitt-Deser-Misner representation of the
higher-derivative gravitational actions}\label{sec3}
We consider the ADM decomposition of the gravitational field
\cite{Arnowitt:1962}. In the first two subsections, however, we shall
work in a more general formalism that does not assume any given basis.
After that the more traditional formalism in ADM coordinate system is
applied. For reviews and mathematical background, see
\cite{ADMreview}.

\subsection{Foliation of spacetime into spatial hypersurfaces}
We consider a globally hyperbolic spacetime $\cM$ that admits a foliation
into a family of nonintersecting Cauchy surfaces $\Sigma_t$, which
cover the spacetime. Each Cauchy surface $\Sigma_t$ is a spacelike
hypersurface, such that every causal curve intersects $\Sigma_t$ exactly
once. These spatial hypersurfaces are parametrized by a global
time function $t$.

The metric tensor $g_{\mu\nu}$ of spacetime induces a metric
$h_{\mu\nu}$ on the spatial hypersurface $\Sigma_t$,
\begin{equation}\label{induced_metric}
h_{\mu\nu} = g_{\mu\nu}+n_\mu n_\nu \,,
\end{equation}
where $n_\mu$ is the future-directed unit normal to $\Sigma_t$.
The metric of spacetime has the signature $(-,+,+,+)$.
Since $n_\mu$ is timelike, it has the norm $n_\mu n^\mu=-1$.
Conversely, the metric of spacetime can be expressed in terms of the
induced metric on $\Sigma_t$ and the unit normal to $\Sigma_t$ as
$g_{\mu\nu} = h_{\mu\nu}-n_\mu n_\nu$.
The induced metric $h_{\mu\nu}$ is sometimes referred to as the first
fundamental form of the hypersurfaces $\Sigma_t$.
With one spacetime index raised, $\projector{\mu}{\nu}
=g^{\mu\rho}h_{\rho\nu}=h^{\mu\rho}h_{\rho\nu}$, it is the projection
operator onto $\Sigma_t$:
\begin{equation}
\projector{\mu}{\nu}=\delta^\mu_\nu+n^\mu n_\nu\,.
\end{equation}
The subscript $\perp$ in front of a tensor is used to denote that it
has been projected onto $\Sigma_t$, thus orthogonal to the normal
$n^\mu$, e.g.,
\begin{equation}
\projected{T}^\mu_{\phantom\mu\nu}=\projector{\mu}{\rho}
\projector{\sigma}{\nu}T^\rho_{\phantom\rho\sigma}\,.
\end{equation}

We denote the metric compatible covariant derivatives on
$(\cM,g_{\mu\nu})$ and $(\Sigma_t,h_{\mu\nu})$ by $\nabla$
and $D$, respectively. The spatial covariant derivative $D$ of a
$(k,l)$-tensor field $T$ on $\Sigma_t$ is given in terms of the
covariant derivative $\nabla$ on spacetime as
\begin{equation}\label{DmuT}
D_\mu
T^{\nu_1\cdots\nu_k}_{\phantom{\nu_1\cdots\nu_k}\rho_1\cdots\rho_l}
= \projector{\sigma}{\mu}
\projector{\nu_1}{\alpha_1}\cdots\projector{\nu_k}{\alpha_k}
\projector{\beta_1}{\rho_1}\cdots\projector{\beta_l}{\rho_l}
\nabla_\sigma
T^{\alpha_1\cdots\alpha_k}_{\phantom{\alpha_1\cdots\alpha_k}
\beta_1\cdots\beta_l} \,,
\end{equation}
where in the right-hand side one considers the extension of $T$ on
spacetime.

The extrinsic curvature tensor of the spatial hypersurface $\Sigma_t$ is
defined as the component of $\nabla_\mu n_\nu$ that is fully tangent to
$\Sigma_t$,
\begin{equation}\label{extrinsic_curvature}
K_{\mu\nu} = \projector{\rho}{\mu}\nabla_\rho n_\nu
= \nabla_\mu n_\nu+n_\mu a_\nu \,,
\end{equation}
where by $a_\mu$ we denote the acceleration of an observer with velocity $n_\mu$, 
\begin{equation}\label{a_mu}
a_\mu = \nabla_n n_\mu = n^\nu\nabla_\nu n_\mu \,.
\end{equation}
The symmetry of $K_{\mu\nu}$ follows from the fact that the shape
operator $u\mapsto\nabla_un$ is self-adjoint, $K(u,v)=\nabla_un\cdot v
=u\cdot\nabla_vn=K(v,u)$, for any vectors $u$ and $v$ tangent to
$\Sigma_t$.
Incidentally, the extrinsic curvature \eqref{extrinsic_curvature} can be
written as the Lie derivative of the induced metric $h_{\mu\nu}$ on
$\Sigma_t$ along the unit normal $n$ to $\Sigma_t$,
\begin{equation}
K_{\mu\nu} = \frac{1}{2}\Lie{n} h_{\mu\nu}\,.
\end{equation}
The trace of the extrinsic curvature is denoted by
$K=h^{\mu\nu}K_{\mu\nu}$.
The extrinsic curvature is sometimes referred to as the second fundamental
form of $\Sigma_t$.

\subsection{Decomposition of curvature tensors with respect to spatial
hypersurfaces}\label{sec3.2}
In order to write the gravitational actions \eqref{S_C} or
\eqref{S_R} in terms of the fundamental forms of the
spatial hypersurfaces $\Sigma_t$ and the unit normal $n_\mu$ to these
hypersurfaces, we have to decompose the curvature tensors into
components tangent and normal to the hypersurfaces.
A detailed account of these standard projection relations is presented,
because one of our projection relations for the Weyl tensor differs from
the ones found in the literature, namely in Ref. \cite{Boulware:1984} and
those following it.

The decomposition of the Riemann tensor of spacetime into components tangent
and normal to the hypersurfaces $\Sigma_t$ is given by the following
projection relations:
\begin{enumerate}[i.]
\item Gauss relation
\begin{equation}\label{GaussEq}
\pR_{\mu\nu\rho\sigma} \equiv
\projector{\alpha}{\mu}\projector{\beta}{\nu}\projector{\gamma}{\rho}
\projector{\delta}{\sigma}R_{\alpha\beta\gamma\delta}
= \sR_{\mu\nu\rho\sigma}  + K_{\mu\rho}K_{\nu\sigma}
- K_{\mu\sigma}K_{\nu\rho} \,;
\end{equation}
\item Codazzi relation
\begin{equation}\label{CodazziEq}
\pR_{\mu\nu\rho\bn} \equiv
\projector{\alpha}{\mu}\projector{\beta}{\nu}\projector{\gamma}{\rho}
n^\delta R_{\alpha\beta\gamma\delta}  = 2D_{[\mu}K_{\nu]\rho} \,;
\end{equation}
\item Ricci relation
\begin{equation}\label{RicciEq}
\pR_{\mu\bn\nu\bn} \equiv
\projector{\alpha}{\mu}n^\beta \projector{\gamma}{\nu}n^\delta
R_{\alpha\beta\gamma\delta} = K_{\mu\rho}K_\nu^{\phantom{\nu}\rho}
- \Lie{n} K_{\mu\nu} + D_{(\mu}a_{\nu)} + a_\mu a_\nu \,.
\end{equation}
\end{enumerate}
The remaining projections of the Riemann tensor are either zero or
related to the given ones by the symmetries of the Riemann tensor.
In the Gauss relation \eqref{GaussEq},
$\sR^{\mu}_{\phantom\mu\nu\rho\sigma}$ is the
Riemann tensor of the three-dimensional hypersurface $\Sigma_t$.
In the used notation, the tensor index $\bn$ has a special meaning,
since it refers to the contraction with the unit normal $n^\mu$. 

For the Ricci tensor $R_{\mu\nu}=R^\rho_{\phantom\rho\mu\rho\nu}$ of
spacetime we obtain the following projection relations 
\begin{align}
\pR_{\mu\nu} &= \sR_{\mu\nu} + K_{\mu\nu}K
- 2K_{\mu\rho}K^\rho_{\phantom{\rho}\nu}
+ \Lie{n}K_{\mu\nu} - D_{(\mu}a_{\nu)} - a_\mu a_\nu
\,,\label{pRmunu}\\
\pR_{\mu\bn} &= D_\nu K^\nu_{\phantom{\nu}\mu} - D_\mu K
\,,\label{Rmun}\\
R_{\bn\bn} &= K_{\mu\nu}K^{\mu\nu} - h^{\mu\nu}\Lie{n}K_{\mu\nu} + D_\mu
a^\mu + a_\mu a^\mu \,,\label{Rnn}
\end{align}
which are obtained from the contractions of the Gauss, Codazzi and Ricci
relations \eqref{GaussEq}--\eqref{RicciEq}.
Note that for any \emph{covariant} tensor $T$ which is tangent to
$\Sigma_t$, its Lie derivative along $n$, $\Lie{n}T$, is also tangent to
$\Sigma_t$. This is because $\Lie{n}\projector{\mu}{\nu}=n^\mu a_\nu$,
and hence $\Lie{n}T$ is equal to its projection on $\Sigma_t$,
$\Lie{n}T=\Lie{n}\projected{T}=\projected{\Lie{n}T}$.

The decomposition of the scalar curvature $R$ of spacetime can be
written as
\begin{equation}\label{R.dec}
\begin{split}
R &=h^{\mu\nu}\pR_{\mu\nu}-R_{\bn\bn}\\
&= \sR + K^2 - 3K_{\mu\nu}K^{\mu\nu} + 2h^{\mu\nu}\Lie{n}K_{\mu\nu} -
2D_\mu a^\mu - 2a_\mu a^\mu\\
&= \sR + K_{\mu\nu}K^{\mu\nu} - K^2 + 2\nabla_\mu \left(n^\mu
K-a^\mu\right),
\end{split}
\end{equation}
where in the last equality we have written the Lie derivative as
\begin{equation}
\Lie{n}K_{\mu\nu} = \nabla_n K_{\mu\nu} + \left(
K_\mu^{\phantom{\mu}\rho} - n_\mu a^\rho \right)
K_{\rho\nu}+
\left( K_\nu^{\phantom{\nu}\rho} - n_\nu a^\rho \right)
K_{\mu\rho}
\end{equation}
in order to write its trace as
\begin{equation}
h^{\mu\nu}\Lie{n}K_{\mu\nu} = \nabla_n K + 2K_{\mu\nu}K^{\mu\nu}
= \nabla_\mu\left(n^\mu K\right) - K^2 + 2K_{\mu\nu}K^{\mu\nu}\,.
\end{equation}
We also used the identity
\begin{equation}
D_\mu a^\mu + a_\mu a^\mu = \nabla_\mu a^\mu \,,
\end{equation}
which can be proven easily by applying \eqref{DmuT} to $D_\mu a^\mu$ and
obtaining the component of $\nabla_\mu a_\nu$ which is fully orthogonal
to $\Sigma_t$, $n^\mu n^\nu\nabla_\mu a_\nu=-a_\mu a^\mu$. For such
decompositions of covariant derivatives of tensors into components
tangent and normal to $\Sigma_t$, see \cite{Chaichian:2012md}.
The last form in \eqref{R.dec} is useful for the \EH part of the action,
since the last term in $\sqrt{-g}R$ is a covariant divergence that can
be written as a surface term. The second form in \eqref{R.dec} is useful
for the curvature-squared part of the action, where the second-order
time derivative terms cannot be written as a divergence.

The Ricci tensor squared is written as a sum of the squares of
its projections \eqref{pRmunu}--\eqref{Rnn}:
\begin{equation}\label{Ricci-squared}
R_{\mu\nu}R^{\mu\nu} = \pR_{\mu\nu}\pR^{\mu\nu}
- 2\pR_{\mu\bn}\pR^\mu_{\phantom{\mu}\bn}
+ \left( R_{\bn\bn} \right)^2 \,.
\end{equation}
The combination of quadratic curvature invariants in the Weyl action
\eqref{S.Weyl} is obtained as
\begin{equation}
\begin{split}
R_{\mu\nu}R^{\mu\nu}-\frac{1}{3}R^2
&= \left( h^{\mu\rho}h^{\nu\sigma}-\frac{1}{3}h^{\mu\nu}h^{\rho\sigma}
\right) \pR_{\mu\nu}\pR_{\rho\sigma}
+\frac{2}{3}R_{\bn\bn} \left( h^{\mu\nu}\pR_{\mu\nu} + R_{\bn\bn}
\right)\\
&\qquad - 2\pR_{\mu\bn}\pR^\mu_{\phantom{\mu}\bn} \,.
\end{split}
\end{equation}

Further, we decompose the Weyl tensor \eqref{Weyl-tensor}
of spacetime into components tangent and normal to the spatial
hypersurfaces $\Sigma_t$. First we obtain the projections of Weyl tensor
where one or two arguments are projected along the unit normal $n$ while
the rest are projected onto $\Sigma_t$:
\begin{equation}\label{pCn}
\begin{split}
\pC_{\mu\nu\rho\bn} &= \pR_{\mu\nu\rho\bn} + \pR_{\bn[\mu}h_{\nu]\rho}
\\
&= 2D_{[\mu}K_{\nu]\rho} + D_\sigma
K^\sigma_{\phantom\sigma[\mu}h_{\nu]\rho} -D_{[\mu}Kh_{\nu]\rho} \\
&= 2\left(
\projector{\alpha}{\mu}\projector{\beta}{\nu}\projector{\gamma}{\rho}
- \projector{\alpha}{[\mu}h_{\nu]\rho} h^{\beta\gamma}
\right) D_{[\alpha}K_{\beta]\gamma}
\end{split}
\end{equation}
and
\begin{equation}\label{pCnn}
\begin{split}
\pC_{\mu\bn\nu\bn} &= \pR_{\mu\bn\nu\bn} +\frac{1}{2}\pR_{\mu\nu}
- \frac{1}{2}h_{\mu\nu}R_{\bn\bn} - \frac{1}{6}h_{\mu\nu}R  \\
&= \frac{1}{2}\left( \projector{\rho}{\mu}\projector{\sigma}{\nu} -
\frac{1}{3}h_{\mu\nu}h^{\rho\sigma} \right) \left( \sR_{\rho\sigma} +
K_{\rho\sigma}K - \Lie{n}K_{\rho\sigma} + D_{(\rho}a_{\sigma)} + a_\rho
a_\sigma \right).
\end{split}
\end{equation}
Finally, we obtain the component of Weyl tensor which is fully tangent
to $\Sigma_t$ as
\begin{equation}\label{pC}
\begin{split}
\pC_{\mu\nu\rho\sigma} &= \pR_{\mu\nu\rho\sigma} -
h_{\mu[\rho}\pR_{\sigma]\nu} + h_{\nu[\rho}\pR_{\sigma]\mu}
+ \frac{1}{3}h_{\mu[\rho}h_{\sigma]\nu}R \\
&= \cK_{\mu\nu\rho\sigma} +h_{\mu\rho}\pC_{\nu\bn\sigma\bn} -
h_{\mu\sigma}\pC_{\nu\bn\rho\bn} -h_{\nu\rho}\pC_{\mu\bn\sigma\bn}
+h_{\nu\sigma}\pC_{\mu\bn\rho\bn} \,,
\end{split}
\end{equation}
where we have defined a new tensor $\cK_{\mu\nu\rho\sigma}$ as
\begin{equation}\label{cK}
\begin{split}
\cK_{\mu\nu\rho\sigma} &= K_{\mu\rho}K_{\nu\sigma} -
K_{\mu\sigma}K_{\nu\rho} - h_{\mu\rho} \left( K_{\nu\sigma}K
-K_{\nu\tau}K^\tau_{\phantom\tau\sigma} \right)
+h_{\mu\sigma} \left( K_{\nu\rho}K
-K_{\nu\tau}K^\tau_{\phantom\tau\rho} \right) \\
&\quad +h_{\nu\rho} \left( K_{\mu\sigma}K
-K_{\mu\tau}K^\tau_{\phantom\tau\sigma} \right)
-h_{\nu\sigma} \left( K_{\mu\rho}K
-K_{\mu\tau}K^\tau_{\phantom\tau\rho} \right) \\
&\quad +\frac{1}{2}\left( h_{\mu\rho}h_{\nu\sigma}
-h_{\mu\sigma}h_{\nu\rho} \right) \left( K^2
-K_{\tau\upsilon}K^{\tau\upsilon} \right).
\end{split}
\end{equation}
This tensor is the traceless part of the quadratic extrinsic curvature
tensor $K_{\mu\rho}K_{\nu\sigma} - K_{\mu\sigma}K_{\nu\rho}$.
Note that $\cK_{\mu\nu\rho\sigma}$ inherits the common symmetries of
Riemann and Weyl tensors.
In \eqref{pC}, we used the fact that in any three-dimensional space, the
Weyl
tensor vanishes necessarily due to its symmetries,
\begin{equation}\label{Weyl-tensor-3d}
\sC_{\mu\nu\rho\sigma}= \sR_{\mu\nu\rho\sigma}  -
2h_{\mu[\rho}\sR_{\sigma]\nu} + 2h_{\nu[\rho}\sR_{\sigma]\mu} +
h_{\mu[\rho}h_{\sigma]\nu}\sR =0\,.
\end{equation}
For this reason the traceless part of \eqref{pC} consists only of the
traceless quadratic extrinsic curvature tensor \eqref{cK}.
Unlike the other projections of Weyl tensor, $\pC_{\mu\nu\rho\sigma}$
is not fully traceless, since it satisfies
\begin{equation}
h^{\nu\sigma}\pC_{\mu\nu\rho\sigma}=\pC_{\mu\bn\rho\bn} \,.
\end{equation}
Evidently $\pC_{\mu\nu\rho\sigma}$ has no trace-trace part, because
$\pC_{\mu\bn\rho\bn}$ is traceless. The
Weyl tensor squared is then obtained as
\footnote{Weyl tensor is expanded in terms of its projections as
\begin{equation*}
\begin{split}
C_{\mu\nu\rho\sigma}&=\pC_{\mu\nu\rho\sigma}
-n_\mu\pC_{\bn\nu\rho\sigma} -n_\nu\pC_{\mu\bn\rho\sigma}
-n_\rho\pC_{\mu\nu\bn\sigma} -n_\sigma\pC_{\mu\nu\rho\bn} \\
&\quad +n_\mu n_\rho\pC_{\bn\nu\bn\sigma}
+n_\mu n_\sigma\pC_{\bn\nu\rho\bn}
+n_\nu n_\rho\pC_{\mu\bn\bn\sigma}
+n_\nu n_\sigma\pC_{\mu\bn\rho\bn} \,.
\end{split}
\end{equation*}
When squared each component of this expansion gives a nonvanishing
contribution only when contracted with itself.}
\begin{equation}\label{Csquared}
C_{\mu\nu\rho\sigma}C^{\mu\nu\rho\sigma} =
\cK_{\mu\nu\rho\sigma}\cK^{\mu\nu\rho\sigma}
+8\pC_{\mu\bn\nu\bn}\pC^{\mu\phantom\bn\nu}_{
\phantom\mu\bn\phantom\nu\bn}
-4\pC_{\mu\nu\rho\bn}\pC^{\mu\nu\rho}_{\phantom{\mu\nu\rho}\bn}\,,
\end{equation}
where as ever the indices of tensors tangent to $\Sigma_t$ are raised
(and lowered) by the induced metric \eqref{induced_metric} on
the hypersurface.

We should emphasize that our result \eqref{pC} for the component of
Weyl tensor which is fully tangent to the spatial hypersurface differs
from the one given in \cite{Boulware:1984}, which has been followed in
the literature to date. In \cite{Boulware:1984}, the component
$\pC_{\mu\nu\rho\sigma}$ is obtained in a form similar to \eqref{pC}
but without the first term $\cK_{\mu\nu\rho\sigma}$.
We obtain that $\pC_{\mu\nu\rho\sigma}$ actually has a nonvanishing
traceless part $\cK_{\mu\nu\rho\sigma}$, in addition to the vanishing
three-dimensional Weyl tensor \eqref{Weyl-tensor-3d}.
In other words, the traceless quadratic extrinsic curvature tensor
defined in \eqref{cK}, which is present in our result \eqref{pC}, is
absent in \cite{Boulware:1984}.

\subsection{ADM variables}
We introduce a timelike vector $t^\mu$ that satisfies
$t^\mu\nabla_\mu t=1$. This vector is decomposed into components
normal and tangent to the spatial hypersurfaces $\Sigma_t$ as
$t^\mu=Nn^\mu+N^\mu$, where $N=-n_\mu t^\mu$ is the lapse function and
$N^\mu=\projector{\mu}{\nu}t^\nu$ is the shift vector on the spatial
hypersurface $\Sigma_t$. The ADM variables consist of the lapse
function, the shift vector and the induced metric \eqref{induced_metric}
on $\Sigma_t$. Together they describe the geometry of spacetime.

Then we introduce a coordinate system on spacetime. We regard the
smooth function $t$ as the time coordinate and introduce an
arbitrary coordinate system ($x^i,i=1,2,3)$ on the spatial
hypersurfaces $\Sigma_t$.
The unit normal to $\Sigma_t$ can now be written in terms of the ADM
variables as
\begin{equation}\label{normal}
n_\mu = -N\nabla_\mu t = (-N,0,0,0) \,,\qquad
n^\mu = \left(\frac{1}{N}, -\frac{N^i}{N}\right) .
\end{equation}
The invariant line element in spacetime is written as
\begin{equation}
ds^2=-N^2dt^2 + h_{ij}(dx^i+N^idt)(dx^j+N^jdt) \,.
\end{equation}
The lapse function must be positive everywhere, $N>0$, since $Ndt$
measures the lapse of proper time between the hypersurfaces $\Sigma_t$
and $\Sigma_{t+dt}$.
In the given ADM coordinate basis, the components of the metric of
spacetime read
\begin{equation}\label{ADMmetric}
g_{00} = -N^2+N_i N^i \,,\qquad  g_{0i} = g_{i0} = N_i
\,,\qquad  g_{ij} = h_{ij} \,,
\end{equation}
where $N_i=h_{ij}N^j$. The contravariant components of the metric of
spacetime are
\begin{equation}\label{ADMinvmetric}
g^{00} = -\frac{1}{N^2} \,,\qquad  g^{0i}= g^{i0} = \frac{N^i}{N^2}
\,,\qquad  g^{ij} = h^{ij}-\frac{N^i N^j}{N^2} \,,
\end{equation}
where $h^{ij}h_{jk}=\delta^i_k$. The indices of tensors that are tangent to
$\Sigma_t$ can be lowered and raised using the induced metric
$h_{ij}$ on $\Sigma_t$ and its inverse $h^{ij}$.

The extrinsic curvature tensor \eqref{extrinsic_curvature} is written as
\begin{equation}\label{K_ij}
K_{ij} = \frac{1}{2}\Lie{n} h_{ij}
= \frac{1}{2N}\left( \p_th_{ij} - 2D_{(i}N_{j)} \right),
\end{equation}
where $\p_t$ denotes the partial derivative with respect to the time
$t$. In the projection relations for the curvature tensors obtained
in Sec.~\ref{sec3.2}, the second-order time derivatives of the metric
$h_{ij}$ are contained in the Lie derivative of the extrinsic curvature,
\begin{equation}\label{LienK}
\Lie{n}K_{ij}=\frac{1}{N}\left( \p_tK_{ij}-\cL_{\vec N}K_{ij} \right),
\end{equation}
where $\cL_{\vec N}$ denotes the Lie derivative along the shift vector
$N^i$ on the spatial hypersurface.

The acceleration \eqref{a_mu} is given by the spatial derivative of the
logarithm of the lapse function as
\begin{equation}
a_\mu = D_\mu\ln N\,.
\end{equation}

In the ADM coordinate basis, the time-components of tensors tangent to
$\Sigma_t$ are defined by the spatial components of the tensor and the
shift vector. For example, $n^\nu A_\mu=n_\mu A^\mu=0$, implies
$A_0=A_iN^i$ and $A^0=0$.

Then we can present the projection relations for the curvature tensors
in terms of ADM variables. The projection relations
\eqref{pRmunu}--\eqref{Rnn} for the Ricci tensor are written
as\footnote{Since we specialize to the ADM coordinates, from now on all
tensors will be tangent to the hypersurface $\Sigma_t$, except the unit
normal $n$. Hence we can omit the prefixed symbol $\perp$ from tensors
that have been projected to the hypersurface, e.g.,
$R_{ij}\equiv\pR_{ij}$.}
\begin{align}
R_{ij} &= \Lie{n}K_{ij} + \sR_{ij} + K_{ij}K
-2K_{ik}K_j^{\phantom{j}k} - \frac{1}{N}D_iD_jN \,,\label{pRij.ADM}\\
R_{i\bn} &= D_jK^j_{\phantom{j}i} - D_iK \,,\label{Rin.ADM}\\
R_{\bn\bn} &= - h^{ij}\Lie{n}K_{ij} + K_{ij}K^{ij} + \frac{1}{N}D^iD_iN
\,,\label{Rnn.ADM}
\end{align}
where $K=h^{ij}K_{ij}$.
The scalar curvature of spacetime \eqref{R.dec} reads 
\begin{equation}\label{R.ADM}
R=2h^{ij}\Lie{n}K_{ij}+\sR+K^2-3K_{ij}K^{ij}-\frac{2}{N}D^iD_i N \,.
\end{equation}
The projection relations \eqref{pCn}--\eqref{pC} for the Weyl tensor are
the following:
\begin{align}
C_{i\bn j\bn} &= -\frac{1}{2}\left( \delta_i^k\delta_j^l -
\frac{1}{3}h_{ij}h^{kl} \right) \left( \Lie{n}K_{kl} - \sR_{kl} -
K_{kl}K
- \frac{1}{N}D_kD_lN \right),\label{pCnn.ADM}\\
C_{ijk\bn} &= 2D_{[i}K_{j]k} + D_lK^l_{\phantom{l}[i}h_{j]k}
-D_{[i}Kh_{j]k}\,,\label{pCn.ADM}\\
C_{ijkl} &= \cK_{ijkl} + h_{ik}C_{j\bn l\bn} - h_{il}C_{j\bn
k\bn} -h_{jk}C_{i\bn l\bn} + h_{jl}C_{i\bn k\bn}\,,\label{pC.ADM}
\end{align}
where the traceless quadratic extrinsic curvature tensor is written as
\begin{equation}\label{cK.ADM}
\begin{split}
\cK_{ijkl} &= K_{ik}K_{jl} - K_{il}K_{jk} -h_{ik} \left( K_{jl}K
-K_{jm}K^m_{\phantom{m}l} \right) +h_{il} \left( K_{jk}K
-K_{jm}K^m_{\phantom{m}k} \right) \\
&\quad +h_{jk} \left( K_{il}K -K_{im}K^m_{\phantom{m}l} \right)
-h_{jl} \left( K_{ik}K -K_{im}K^m_{\phantom{m}k} \right) \\
&\quad + \frac{1}{2}\left( h_{ik}h_{jl}-h_{il}h_{jk} \right)
\left( K^2 -K_{mn}K^{mn} \right).
\end{split}
\end{equation}

The Weyl tensor squared \eqref{Csquared} is obtained in the form
\begin{equation}\label{Csquared.ADM}
C_{\mu\nu\rho\sigma}C^{\mu\nu\rho\sigma} =
8C_{i\bn j\bn}C^{i\phantom\bn j}_{\phantom i\bn\phantom j\bn}
-4C_{ijk\bn}C^{ijk}_{\phantom{ijk}\bn}\,,
\end{equation}
since $\cK_{ijkl}$ squared is zero due to the Cayley-Hamilton theorem.
We indeed obtain
\begin{equation}
\cK_{ikjl}K^{kl}=-3P(K)_{ij}=0
\end{equation}
and consequently
\begin{equation}\label{cKsquared.ADM}
\cK_{ijkl}\cK^{ijkl}
=-6P(K)^i_{\phantom ij}K^j_{\phantom ji}=0 \,,
\end{equation}
where $P(K)^i_{\phantom ij}$ is the characteristic
polynomial\footnote{See Appendix~\ref{appendix2} for the Cayley-Hamilton
theorem and the definition of the characteristic polynomial.},
\eqref{P(A)=0.3d}, for the tensor $K^i_{\phantom ij}=h^{ik}K_{kj}$
with the tensor itself as the argument, which is identically zero.
This means that the correction to the projection relation
\eqref{pC.ADM} -- namely the tensor \eqref{cK.ADM} -- has no impact on
the Hamiltonian formulation of the given gravitational theories
\eqref{S_C}. If the Weyl tensor were coupled to a tensor (or
tensors) other than itself, the contribution of $\cK_{ijkl}$ would not
vanish in general.

\subsection{First-order ADM representation of the action}\label{sec3.4}
Let us consider the gravitational actions \eqref{S_C} and
\eqref{S_R} without the \GBC topological invariant term proportional to $\int
d^4x\sqrt{-g}G$, i.e. we set $\gamma=0$ in \eqref{S_C}, and $\gamma-\alpha/4=0$ in \eqref{S_R} .
We shall present the actions in terms of the foliation of spacetime
into spatial hypersurfaces $\Sigma_t$, using the ADM variables and an
associated coordinate system.

Since the invariant terms in the action which are quadratic in curvature
contain second-order time derivatives \eqref{LienK}, we are dealing with
a higher-derivative theory. In the Lagrangian formalism this is not a
problem at all, because the Euler-Lagrange equations can in principle
contain any number of time derivatives, in the present case up to
fourth order. In the Hamiltonian formalism, the equations of motion
contain only a first-order time derivative, namely in the form
$df/dt=\pb{f,H}$.
In order to achieve such a first-order description of the dynamics of
the higher-derivative action, we shall introduce additional variables
and constraints so that the action can be rewritten into a form which
contains only first-order time derivatives. The additional variables
describe the extra degrees of freedom implied by the higher-order time
derivatives. In the present case, it is most convenient to regard the
metric $h_{ij}$ and the extrinsic curvature $K_{ij}$ as independent
variables. The fact that $h_{ij}$ and $K_{ij}$ are related is taken
into account by imposing their relation \eqref{K_ij} as a constraint,
using Lagrange multipliers. Thus from now on we consider the
higher-derivative gravitational Lagrangian as a functional of the
independent variables $N$, $N^i$, $h_{ij}$ and $K_{ij}$. It also depends on the
first-order time derivative of $K_{ij}$, and we extend it with the
constraint $\Lie{n}h_{ij}-2K_{ij}=0$ multiplied by the Lagrange multiplier
$\lambda^{ij}$. That is we understand the complete action
(without matter) as a functional
\begin{equation}\label{S.second-order}
S[N,N^i,h_{ij},\dot{h}_{ij},K_{ij},\dot{K}_{ij},\lambda^{ij}]\,,
\end{equation}
where the dot denotes time derivative.\footnote{There is some room for
the choice of what is regarded as the time derivative in Hamiltonian
formulation of the theory. This is discussed in Sec.~\ref{sec4}.}
Now the extra degrees of freedom associated with the second-order time
derivative of the metric are carried by the variables $K_{ij}$.

The first-order ADM representation of the action \eqref{S_C}
can now be written as
\begin{multline}\label{S_C.dec}
S_C = \int dt\int_{\Sigma_t}d^3x N\sqrt{h}\left[ \Lambda +
\frac{1}{2\kappa}\left( \sR+K_{ij}K^{ij}-K^2 \right)
-2\alpha C_{i\bn j\bn}C^{i\phantom\bn j}_{\phantom{i}
\bn\phantom{j}\bn} \right.\\
+\left. \alpha C_{ijk\bn}C^{ijk}_{\phantom{ijk}\bn}
+\frac{\beta}{8}R^2
+\lambda^{ij}\left( \Lie{n}h_{ij}-2K_{ij} \right) \right]
 + \Ssurface \,,
\end{multline}
where the expressions \eqref{R.ADM}--\eqref{pCn.ADM} are assumed and 
$\Ssurface$ contains the surface terms.
Alternatively, we could use the action \eqref{S_R}, whose first-order
ADM representation is expressed as
\begin{multline}\label{S_R.dec}
S_R = \int dt\int_{\Sigma_t}d^3x N\sqrt{h}\left[ \Lambda +
\frac{1}{2\kappa}\left(
\sR+K_{ij}K^{ij}-K^2 \right)
- \frac{\alpha}{2}\left( h^{ik}h^{jl}-\frac{1}{3}h^{ij}h^{kl} \right)
R_{ij}R_{kl} \right.\\
-\left.\frac{\alpha}{3}R_{\bn\bn} \left( \sR+K^2-K_{ij}K^{ij} \right)
+\alpha R_{i\bn}R^i_{\phantom{i}\bn}
+\frac{\beta}{8}R^2
+\lambda^{ij}\left( \Lie{n}h_{ij}-2K_{ij} \right) \right]\\
+ \Ssurface \,,
\end{multline}
and we assumed \eqref{pRij.ADM}--\eqref{R.ADM}.

These two forms of the action \eqref{S_C.dec} and \eqref{S_R.dec} differ, as mentioned previously,
only by a multiple of the \GBC topological invariant term,
$-(\alpha/4)\int d^4x\sqrt{-g}G$, albeit this is no longer so evident
because it too has been decomposed with respect to the spatial
hypersurfaces as a part of the Weyl tensor squared term
\eqref{Weyl.squared}. First of all the kinetic part of the Lagrangian
density of the action \eqref{S_C.dec} is simpler than the one of
\eqref{S_R.dec}. In particular, the Weyl gravity part of the Lagrangian
density of \eqref{S_C.dec} has no dependence on $h^{ij}\Lie{n}K_{ij}$,
whereas the Lagrangian density of \eqref{S_R.dec} contains the linear
term $\frac{\alpha}{3}h^{ij}\Lie{n}K_{ij} \left( \sR+K^2-K_{kl}K^{kl}
\right)$. This difference has consequences for the
structure of the constraints in the Hamiltonian formulations of these
two forms of the gravitational action.

\subsubsection{Surface terms}
In the case of general relativity, the surface terms are obtained as
\begin{equation}\label{S_surface.EH}
\Ssurface= \frac{1}{\kappa}\oint_{\p\cM}d^3x\sqrt{|\gamma|}K
+\frac{1}{\kappa}\oint_{\p\cM}d^3x\sqrt{|\gamma|}r_\mu \left(n^\mu
K-a^\mu\right),
\end{equation}
where $\gamma_{\mu\nu}$ and $r_\mu$ are the induced metric and the
outward-pointing unit normal to the boundary $\p\cM$ of spacetime,
respectively. We should emphasize that in the first surface term, $K$
refers to the extrinsic curvature of the boundary $\p\cM$, while in the
last term, $K$ refers to the extrinsic curvature of the spatial
hypersurfaces $\Sigma_t$.
In our globally hyperbolic spacetime, the boundary $\p\cM$
consists of the initial and final Cauchy surfaces, say $\Sigma_0$ and
$\Sigma_1$, respectively, and of the timelike hypersurface $\cB$ that
connects those spatial hypersurfaces. The timelike part of the boundary
is the union $\cB=\bigcup_{t\in\mathbb{R}}\cB_t$ of the two-dimensional
boundaries $\cB_t$ of the Cauchy surfaces $\Sigma_t$ (at spatial
infinity). On the initial and final Cauchy surfaces $\Sigma_0$ and
$\Sigma_1$, the surface integrals cancel each other entirely. Thus only
the integral over $\cB$ survives in the surface terms,
\begin{equation}\label{S_surface}
\Ssurface= \frac{1}{\kappa}\int_{\cB} d^3x\sqrt{-\gamma}\left(
K_\cB + r_\mu n^\mu K - r_\mu a^\mu \right).
\end{equation}
Here the trace of the extrinsic curvature of $\cB$ is denoted by
$K_\cB=\nabla_\mu r^\mu$, so that it is not confused with $K$ which
is to the trace of the extrinsic curvature of the surfaces
$\Sigma_t$ on its intersection with the boundary $\cB$.
If the surfaces $\cB$ and $\Sigma_t$ are assumed to be orthogonal,
the normals to $\cB$ and $\Sigma_t$ are orthogonal as well, i.e., $r_\mu
n^\mu=0$, and hence we further obtain \cite{Hawking:1995fd}
\begin{equation}\label{S_surface.EH.ort}
\Ssurface= \frac{1}{\kappa}\int_{\cB} d^3x\sqrt{-\gamma}
h^{\mu\nu}\nabla_\mu r_\nu
= \frac{1}{\kappa}\int dt\oint_{\cB_t}d^2x\sqrt{\sigma}\,N\Kt \,,
\end{equation}
where $\sigma_{ab}$ is the induced metric on $\cB_t$ and $\Kt$ is
the extrinsic curvature of $\cB_t$ embedded in $\Sigma_t$. In the case
of nonorthogonal boundaries, one actually has to include extra
two-dimensional surface terms regarding the intersection angle
$\eta=r_\mu n^\mu$ of $\cB$ and $\Sigma_t$ as \cite{Hawking:1996ww}
\begin{equation}\label{S_surface.EH.nonort}
\Ssurface=  \frac{1}{\kappa}\int_{\cB} d^3x\sqrt{-\gamma}\left(
K_\cB + \eta K - r_\mu a^\mu \right)
+\frac{1}{\kappa}\int^{\cB_1}_{\cB_0}d^2x\sqrt{\sigma}\sinh^{-1}\eta
 \,,
\end{equation}
where we denote the difference of the integrals over the two-dimensional
final and initial surfaces $\cB_1$ and $\cB_0$ as
$\int^{\cB_1}_{\cB_0}=\int_{\cB_1}-\int_{\cB_0}$.

As was noted in Sec.~\eqref{sec2.3}, if the spacetime is spatially
noncompact, we must choose a reference background and define the
physical action as the difference to the reference action.
This also applies to all surface terms in the action

Let us then consider the variational principle for the higher-derivative
gravitational action \eqref{S.second-order}. Now the action depends on both the induced metric $h_{ij}$ and extrinsic
curvature $K_{ij}$ on the hypersurface $\Sigma_t$, viewed as independent variables. Therefore, we should
consider variations of the action for which both $h_{ij}$ and $K_{ij}$
are held fixed on the boundary of spacetime.
Thus we require that the solutions to the equations of motion \eqref{EoM}
lead to extrema of the action when the variations of all the variables are
imposed to be zero on the boundary $\p\cM$.
Now consider the surface terms obtained in Sec.~\ref{sec2.3}. The
surface terms \eqref{EH-surfaceterm} and \eqref{Rsquared-surfaceterm}
obtained from the \EH and scalar curvature squared parts of the action
clearly vanish under the variations with respect to the enlarged
configuration space of the action \eqref{S.second-order}, because now
the variation $\delta K$ of the trace of extrinsic curvature vanishes on
the boundary $\p\cM$. The surface term \eqref{Weyl-surfaceterm} which
is implied by the variation of the action of Weyl gravity is a more
complicated matter. By decomposing the integrand of this surface term
into components tangent and normal to the boundary surface, it can be
shown that the integrand is linear in the variations of the ADM
variables and the extrinsic curvature. (See \cite{Barth:1985} for the
details of the calculation.) Thus the surface term
\eqref{Weyl-surfaceterm} also is zero when the variations of the ADM
variables and the extrinsic curvature are imposed to vanish on the
boundary. Boundary terms in curvature-squared gravity have also been
studied in \cite{Horowitz:1985} with the same result: boundary terms are
no longer required when quadratic curvature invariants are added into
the \EH action. Recently, the same conclusion was reached in
\cite{Grumiller:2013mxa}.
Therefore the only surface term at this point is the one that
originates from the covariant total derivative in the decomposition
of the scalar curvature \eqref{R.dec} in the \EH action, namely
\begin{equation}\label{R-surfaceterm}
\frac{1}{\kappa}\oint_{\p\cM}d^3x\sqrt{|\gamma|}r_\mu
\left(n^\mu K-a^\mu\right).
\end{equation}
Note that this surface term could be easily avoided by using the second
expression for the decomposition of the scalar curvature of spacetime
\eqref{R.dec} and keeping the time derivative of $K_{ij}$. We, however,
prefer to use the last expression of \eqref{R.dec}, and hence obtain the
surface term \eqref{R-surfaceterm} whenever the \EH action is present.
The fact that no boundary terms are required in curvature-squared
gravity does not mean that it is forbidden to include boundary terms
into the gravitational action. Indeed, we can include any boundary term
whose variation is linear in the variations of the ADM variables and the
extrinsic curvature. Such boundary terms do not compromise the action
principle due to the chosen boundary conditions on the variations.
Whenever the \EH action is present, we shall take advantage of this
freedom by including the same boundary term that is required in general
relativity, $\kappa^{-1}\oint_{\p\cM}d^3x\sqrt{|\gamma|}K$, so that
combined with \eqref{R-surfaceterm}, the total surface term takes the
same form as in general relativity \eqref{S_surface.EH}: more
specifically \eqref{S_surface.EH.ort}, when the hypersurfaces are
orthogonal or \eqref{S_surface.EH.nonort}, if the hypersurfaces are
nonorthogonal. This choice is motivated by the fact that the surface
term plays the role of  total energy in the Hamiltonian formulation, and
we prefer to obtain a similar total energy as in general relativity.
(See \cite{Hawking:1995fd} for the case of general relativity.)
In pure curvature-squared gravity, when the \EH action is absent, we
shall not include any surface terms, $\Ssurface=0$.

\section{Hamiltonian analysis}\label{sec4}
The Hamiltonian formulation and canonical quantization of gravitational
theories whose Lagrangians are quadratic in curvature were
originally studied in \cite{Kaku:1983,Boulware:1984,Buchbinder:1987}.
These Hamiltonian formulations differ significantly. Kaku
\cite{Kaku:1983} formulated conformally invariant Weyl gravity in the
strong-coupling approximation, both in Hamiltonian and Lagrangian forms,
analogous to higher-derivative Yang-Mills theory. The differences of
\cite{Boulware:1984} and \cite{Buchbinder:1987} are particularly
interesting. Boulware \cite{Boulware:1984} based his analysis on the
action \eqref{S_C} without the topological invariant term $\gamma\int
d^4x\sqrt{-g}G$, while Buchbinder and Lyahovich \cite{Buchbinder:1987}
considered an action of the form \eqref{S_R} without the topological
invariant term. In Sec.~\ref{sec3}, we obtained first-order forms
of both of these actions in terms of ADM variables. They are given in
\eqref{S_C.dec} and \eqref{S_R.dec}, respectively. We shall choose the
action \eqref{S_C.dec} as the basis of our Hamiltonian formulation of
curvature-squared gravity, because of its simpler kinetic part compared
to the action \eqref{S_R.dec}. The Hamiltonian analysis based on the
action \eqref{S_R.dec} would indeed result into more complicated
constraints, even if one  uses a canonical transformation for its
simplification \cite{Buchbinder:1987}. Those complications in the
structure of constraints will be remarked upon in the following
analysis.

There are a few plausible choices for what is regarded as the time
derivative in the Hamiltonian formulation of the given higher-derivative
gravitational theory. The obvious and most common choice is
to consider the partial derivative with respect to time as the concept
of time differentiation for Hamiltonian formulation of gravity,
following the original ADM formalism \cite{Arnowitt:1962}.
 However, in principle, we could choose the derivative along any
(nondynamical) timelike vector as a generalized time derivative
for Hamiltonian formulation. Since the given gravitational Lagrangian
of \eqref{S_C.dec} or \eqref{S_R.dec} is independent of the time
derivatives of $N$ and $N^i$, and we know from previous analyses that
they behave as arbitrary Lagrange multipliers, the unit normal
$n^\mu$ to the spacelike hypersurface is not a dynamical quantity.
Furthermore, time derivatives in the actions appear only in the form of
$\Lie{n}h_{ij}$ and $\Lie{n}K_{ij}$, thus making the Lie derivative
$\Lie{n}$ along the unit normal $n^\mu$ a tempting alternative for the
concept of time differentiation for the Hamiltonian
formulation.\footnote{Another alternative for the concept of time
differentiation would be the Lie derivative $\Lie{m}$, where the vector
$m^\mu=Nn^\mu=(1,-N^i)$ is the component of the time vector $t^\mu$
which is normal to the hypersurfaces $\Sigma_t$. Then we would have
$\Lie{n}h_{ij}=N^{-1}\Lie{m}h_{ij}$ and
$\Lie{n}K_{ij}=N^{-1}\Lie{m}K_{ij}$.}
This kind of approach was adopted in \cite{Boulware:1984}, and later followed in \cite{Demaret:1995,Querella:1998}.
On the other hand, the nondynamical nature of $N$ and $N^i$ is a result
of the Hamiltonian analysis, rather than its premise, because it is not
evident from the beginning that $N$ and $N^i$ do not appear in any of
the constraints of the theory in Hamiltonian formulation. 
We shall treat $N$ and $N^i$ as genuine variables in the Hamiltonian
analysis, uncovering that they can be
considered as Lagrange multipliers.
We consider partial derivative with respect to time ($\p_t$) as
the concept of time differentiation in the following Hamiltonian
formulation of the theory.

The Lagrangian density in the action \eqref{S_C.dec} is a function
\begin{equation}\label{L_C.function}
\cL_C(N,N^i,h_{ij},\p_th_{ij},K_{ij},\p_tK_{ij},\lambda^{ij})\,.
\end{equation}
The canonical momenta conjugated to $N$ and $N^i$ are primary
constraints
\begin{equation}\label{pconstraints.1}
p_N\approx0\,,\qquad p_i\approx0\,,
\end{equation}
respectively, since the action is independent of the time derivatives of
$N$ and $N^i$. The weak equality ($\approx$) is understood in the sense
introduced by Dirac \cite{Dirac:Lectures1964}: a weak equality can be
imposed only after all Poisson brackets have been evaluated, while a
usual strong equality could be imposed anywhere.
The tensor density defined by the Lagrange multiplier $\lambda^{ij}$
is identified as the canonical momentum conjugated to $h_{ij}$,
\begin{equation}\label{p^ij}
p^{ij}=\frac{\p\cL_C}{\p(\p_th_{ij})}=\sqrt{h}\lambda^{ij}\,.
\end{equation}
The canonical momentum conjugated to $K_{ij}$ is defined as
\begin{equation}\label{cP^ij}
\cP^{ij}=\frac{\p\cL_C}{\p(\p_tK_{ij})}=\sqrt{h} \left( 2\alpha
C^{i\phantom\bn j}_{\phantom{i}\bn\phantom{j}\bn}
+\frac{\beta}{2}h^{ij}R \right).
\end{equation}

Note that once we have identified $\sqrt{h}\lambda^{ij}$ as the
canonical momentum $p^{ij}$ conjugate to $h_{ij}$ in \eqref{p^ij},
it is unnecessary to include the Lagrange multiplier $\lambda^{ij}$ and
its conjugated momentum $p^\lambda_{ij}$ as extra canonical variables.
If we include them, we obtain the extra primary constraints
$p^{ij}-\sqrt{h}\lambda^{ij}\approx0$ and $p^\lambda_{ij}\approx0$. We
can set these second-class constraints to zero strongly and eliminate
the variables $\lambda^{ij}$ and $p^\lambda_{ij}$ by substituting
$\sqrt{h}\lambda^{ij}=p^{ij}$ and $p^\lambda_{ij}=0$. The Dirac bracket
defined by these second-class constraints is equivalent to the Poisson
bracket for all the remaining variables (see Appendix~\ref{appendix3}
for details). In general, this applies to any higher-derivative theory,
where the relevant primary constraints are linear in the Lagrange
multipliers.

The number and nature of constraints and physical degrees of freedom
depends on the values of the coupling constants. Therefore we shall
treat the different cases separately. We are particularly interested in
the cases with $\alpha\neq0$, which possess the potential to be
renormalizable, that is consistent at high energies. The cases
with $\alpha=0$ include only general relativity and a special case
of $f(R)$ gravity, $f(R)=R+bR^2$, with or without the cosmological
constant term, which are well known and understood.
First we shall consider the most interesting case, namely the
conformally invariant Weyl gravity. Weyl gravity will serve as the
reference theory to which all the other cases are compared.

\subsection{Weyl gravity\texorpdfstring{: $\Lambda=0$,
$\kappa^{-1}=\beta=\gamma=0$, $\alpha\neq0$}{}}\label{sec4.1}
First we consider the case of conformally invariant Weyl gravity
\eqref{S.Weyl}. The action is given in \eqref{S_C.dec} with $\Lambda=0$
and the couplings $\kappa^{-1}=\beta=0$, $\alpha\neq0$, and without any
surface terms $\Ssurface=0$.  We could  also set the coupling
$\alpha=1$, but we choose not to, because keeping it will help in
comparing to the other cases. The topological invariant term in
\eqref{S_C} has been discarded, $\gamma=0$.

The momentum \eqref{cP^ij} consists only of the projection
\eqref{pCnn.ADM} of the Weyl tensor. It can be written as
\begin{equation}\label{cP^ij.Weyl}
\cP^{ij}=-\alpha\sqrt{h}\bcG^{ijkl}\left( \Lie{n}K_{kl} - \sR_{kl} -
K_{kl}K - \frac{1}{N}D_kD_lN \right),
\end{equation}
where we have defined a traceless generalized DeWitt metric as
\begin{equation}\label{bcG^ijkl}
\bcG^{ijkl}=\frac{1}{2}\left( h^{ik}h^{jl}+h^{il}h^{jk} \right)
-\frac{1}{3}h^{ij}h^{kl}\,.
\end{equation}
Since the Weyl tensor is traceless, in other words the DeWitt metric has
the metric as its null eigenvector, $g_{ij}\bcG^{ijkl}=0
=\bcG^{ijkl}g_{kl}$, the trace of the momentum \eqref{cP^ij.Weyl} is
zero. Thus we have to define one more primary constraint
\footnote{If we based our Hamiltonian formulation on the action
\eqref{S_R.dec}, instead of \eqref{S_C.dec}, this constraint would have
the form
\[
\cP-\alpha\sqrt{h}\left( \sR+K^2-K_{ij}K^{ij} \right)\approx0\,,
\]
where the extra terms compared to \eqref{P.constraint} depend on both
$h_{ij}$ and $K_{ij}$. These extra terms would complicate the analysis
significantly.}
\begin{equation}\label{P.constraint}
\cP\approx0\,.
\end{equation}
We adopt the notation where the trace of a quantity is denoted without
indices and the traceless part is denoted by the bar accent. For
example, we denote
\begin{equation}
\cP=h_{ij}\cP^{ij}\,,\qquad \bcP^{ij}=\cP^{ij}-\frac{1}{3}h^{ij}\cP\,.
\end{equation}
The DeWitt metric \eqref{bcG^ijkl} has the traceless inverse
\begin{equation}\label{bcG_ijkl}
\bcG_{ijkl}=\frac{1}{2}\left( h_{ik}h_{jl}+h_{il}h_{jk} \right)
-\frac{1}{3}h_{ij}h_{kl}\,,
\end{equation}
which satisfies
\begin{equation}
\bcG_{ijmn}\bcG^{mnkl}=
\delta_i^{(k}\delta_j^{l)}-\frac{1}{3}h_{ij}h^{kl}\,.
\end{equation}
The definition of the momentum \eqref{cP^ij.Weyl} can then be used to
obtain
\begin{equation}
\begin{split}
\cP^{ij}\p_tK_{ij}&=-N\frac{\bcP_{ij}\bcP^{ij}}{\alpha\sqrt{h}}
+N\bcP^{ij} \left( \sR_{ij} +K_{ij}K \right) +\bcP^{ij}D_iD_jN \\
&\quad +\cP^{ij}\cL_{\vec N}K_{ij}+\frac{N}{3}\cP h^{ij}\Lie{n}K_{ij}
\,,
\end{split}
\end{equation}
where $\bcP_{ij}=\bcG_{ijkl}\cP^{kl}$.
The Lagrangian density of the action is written in terms of the
canonical variables as
\begin{equation}
\begin{split}
\cL_C &= -N\left( \frac{\bcP_{ij}\bcP^{ij}}{2\alpha\sqrt{h}}
+2p^{ij}K_{ij} -\alpha\sqrt{h}C_{ijk\bn}C^{ijk}_{\phantom{ijk}\bn}
\right)  +p^{ij}\p_th_{ij} -2p^{ij}D_{(i}N_{j)}\,.
\end{split}
\end{equation}

By definition, the total Hamiltonian is
\begin{equation}\label{H.Weyl}
\begin{split}
H &= \int_{\Sigma_t}d^3x\left( p^{ij}\p_th_{ij} + \cP^{ij}\p_tK_{ij}
- \cL_C + u_N p_N + u^i p_i  + u_\cP\cP \right) \\
&= \int_{\Sigma_t}d^3x\left( N\cH_0 + N^i\cH_i + \lambda_N p_N
+ \lambda^ip_i + \lambda_\cP\cP \right) + \Hsurface,
\end{split}
\end{equation}
where all the $u$ and $\lambda$ are arbitrary Lagrange
multipliers accounting for the first two constraints in \eqref{pconstraints.1}, as well as the constraint \eqref{P.constraint}. In the Hamiltonian, we have defined the following
quantities. The Hamiltonian constraint is given as
\begin{equation}\label{H_0.Weyl}
\cH_0 = 2p^{ij}K_{ij} -\frac{\cP_{ij}\cP^{ij}}{2\alpha\sqrt{h}}
+\cP^{ij}\sR_{ij} +\cP^{ij}K_{ij}K +D_iD_j\cP^{ij}
-\alpha\sqrt{h}C_{ijk\bn}C^{ijk}_{\phantom{ijk}\bn} \,,
\end{equation}
where $\cP_{ij}=h_{ik}h_{jl}\cP^{kl}$.
We have written the Hamiltonian constraint in terms of all the
components of the momentum $\cP^{ij}$, absorbing terms depending on
the trace component $\cP$ into the Lagrange multiplier term
$\lambda_\cP\cP$. The momentum constraint has the form
\begin{equation}\label{H_i.Weyl}
\cH_i=-2h_{ij}D_kp^{jk} +\cP^{jk}D_iK_{jk} -2D_j\left( \cP^{jk}K_{ik}
\right)\,,
\end{equation}
or, in terms of partial derivatives,
\begin{equation}
\cH_i=-2h_{ij}\p_kp^{jk} -\left( 2\p_jh_{ik}-\p_ih_{jk} \right)p^{jk}
-2K_{ij}\p_k\cP^{jk} -\left( 2\p_jK_{ik}-\p_iK_{jk} \right)\cP^{jk} \,.
\end{equation}
The surface term in the Hamiltonian \eqref{H.Weyl} is expressed as
\begin{equation}
\Hsurface= \oint_{\cB_t}d^2x s_i\left( D_jN\cP^{ij}-ND_j\cP^{ij}
+2N_jp^{ij}+2N^j\cP^{ik}K_{jk} \right),
\end{equation}
where $s_i$ is the unit normal to the spatial boundary $\cB_t$ embedded
in $\Sigma_t$.
The surface terms appear for two reasons. The first two appear due to the
integration by parts of the term
\begin{equation*}
\int_{\Sigma_t}d^3x\cP^{ij}D_iD_jN = \int_{\Sigma_t}d^3xND_iD_j\cP^{ij}
+ \oint_{\cB_t}d^2x s_i\left( D_jN\cP^{ij}-ND_j\cP^{ij} \right).
\end{equation*}
The last two surface terms come from the integration by parts of the
momentum constraint. Indeed, we define a smeared momentum
constraint as the functional
\begin{equation}\label{MC}
\MC{\vec X}=\int_{\Sigma_t}d^3x X^i\cH_i\,,
\end{equation}
where $\vec X$ is an arbitrary test vector on $\Sigma_t$.
The momentum constraint \eqref{MC} can be written as
\begin{equation}
\MC{\vec X}=\int_{\Sigma_t}d^3x\left( p^{ij}\cL_{\vec X}h_{ij}
+\cP^{ij}\cL_{\vec X}K_{ij} \right) - \oint_{\cB_t}d^2x 2s_i\left(
X_jp^{ij}+X^j\cP^{ik}K_{ik} \right),
\end{equation}
where $\cL_{\vec X}h_{ij}=2D_{(i}X_{j)}$, which shows the origin of the
last two surface terms.
Thus the momentum constraint evidently generates infinitesimal
(time-dependent) spatial diffeomorphism for the dynamical variables
$(h_{ij},p^{ij},K_{ij},\cP^{ij})$ on the hypersurface $\Sigma_t$.
We can extend the momentum constraint to a generator of
(time-dependent) spatial diffeomorphism for all variables
by absorbing certain terms into the Lagrange multipliers of the primary
constraints \eqref{pconstraints.1}. It can be defined up to boundary
terms as
\begin{equation}
\MC{\vec X}=\int_{\Sigma_t}d^3x\left( p^{ij}\Lie{\vec X}h_{ij}
+\cP^{ij}\Lie{\vec X}K_{ij} +p_N\Lie{\vec X}N +p_i\Lie{\vec X}N^i
 \right).
\end{equation}
In either case, the momentum constraint satisfies the Lie algebra
\begin{equation}\label{MC-Liealgebra}
\pb{\MC{\vec X},\MC{\vec Y}}=\MC{\bigl[\vec X,\vec Y\bigr]}\,,
\end{equation}
due to the corresponding property of the Lie derivative,
\begin{equation}
\Lie{\vec X}\Lie{\vec Y}-\Lie{\vec Y}\Lie{\vec X}=\Lie{[\vec X,\vec Y]}
\,,\qquad \bigl[\vec X,\vec Y\bigr]^i=X^j\p_jY^i-Y^j\p_jX^i\,.
\end{equation}
The variables $N$, $N^i$, $h_{ij}$ and $K_{ij}$ behave as regular
scalar or tensor fields under the spatial diffeomorphisms, while their
canonically conjugated momenta behave as scalar or tensor densities of
unit weight. Thus we can see that all the constraints behave as scalar
or tensor densities of unit weight under the spatial diffeomorphisms.

We also define a smeared version of the Hamiltonian constraint as
\begin{equation}\label{HC}
\HC{\xi}=\int_{\Sigma_t}d^3x \xi\cH_0\,,
\end{equation}
where $\xi$ is an arbitrary test function on $\Sigma_t$.
It satisfies the following algebra with the momentum constraint
\begin{equation}
\pb{\MC{\vec X},\HC{\xi}}=\HC{\vec X(\xi)} \,,
\end{equation}
since $\cH_0$ is a scalar density of unit weight and consequently it
satisfies
\begin{equation}
\pb{\cH_0,\MC{\vec X}}=\Lie{\vec X}\cH_0=X^i\p_i\cH_0+\p_iX^i\cH_0 \,.
\end{equation}

Note that in the Hamiltonian, we could alternatively replace $\cP^{ij}$ 
with its traceless components $\bcP^{ij}$, or vice versa,
because any term depending on a positive power of the primary constraint
$\cP$ can be absorbed into the Lagrange multiplier term
$\lambda_\cP\cP$. Thus we could equally well define $\cH_0$ in
\eqref{H_0.Weyl} with every $\cP^{ij}$ replaced by $\bcP^{ij}$. We
shall, however, write the Hamiltonian in terms of all the components of
the momentum $\cP^{ij}$, since it simplifies slightly the calculation of
the Poisson brackets between the constraints.
The same applies to the momentum constraint and the surface terms.
Note that we have written the first two surface terms in \eqref{H.Weyl}
with the full momentum $\cP^{ij}$, corresponding to the term
$D_iD_j\cP^{ij}$ in \eqref{H_0.Weyl}. The momentum constraint
\eqref{H_i.Weyl} too is written with all the momenta so that it
generates diffeomorphisms also for the trace component $\cP$.
Hence the last surface term in the Hamiltonian involves all the
components of $\cP^{ij}$ as well.

\subsubsection{Consistency of constraints in time and 
secondary constraints}
Every constraint has to be preserved under time evolution. This means
the algebra of constraints has to be closed under the Poisson bracket.

The consistency of the primary constraints $p_N$ and $p_i$ in time is
ensured by imposing $\cH_0$ and $\cH_i$ as local constraints,
\begin{equation}
\cH_0\approx0\,,\qquad \cH_i\approx0\,,
\end{equation}
respectively. This is why they were above referred to as the
Hamiltonian constraint and the momentum constraint, respectively.
The momentum constraint means that the theory is invariant under
diffeomorphisms on the spatial hypersurface, i.e., generally covariant.
The Hamiltonian constraint contains the dynamics of the theory.
These constraints (at every point of $\Sigma_t$) are independent
restrictions on the canonical variables \cite{Teitelboim:1973zz}.
Because the time evolution of the lapse $N$ and shift $N^i$
variables is given by the arbitrary Lagrange multipliers $\lambda_N$ and
$\lambda^i$, the lapse and shift variables themselves behave indeed
as arbitrary multipliers in the Hamiltonian.

The consistency condition for the primary constraint  $\cP$ implies a
secondary constraint. We express this new constraint as
\footnote{If we based our Hamiltonian formulation on the action
\eqref{S_R.dec}, this constraint would contain an extra term of the
form $\alpha\sqrt{h}\left( D^iD^jK_{ij}-D^iD_iK \right)$ (up to a
numerical factor). This extra term would further complicate the analysis
significantly.}
\begin{equation}\label{Q.constraint}
\cQ=2p+\cP^{ij}K_{ij}\approx0\,.
\end{equation}
Note that we have included the trace component $\cP\approx0$ in the
constraint $\cQ$, similarly as we did in the Hamiltonian and momentum
constraints, and for the same reason.
Thanks to the secondary constraint \eqref{Q.constraint}, $\cP$ is
preserved in time:
\begin{equation}
\pb{\cP(\bx),H}\approx \pb{\cP(\bx),\HC{N}}
=-N\left( \cQ +\cP K \right)(\bx) \approx0\,.
\end{equation}
The Poisson bracket between $\cP$ and $\cQ$ closes,
\begin{equation}
\pb{\cP(\bx),\cQ(\by)}=\cP(\by)\delta(\bx-\by)\,.
\end{equation}
Then we have to ensure that the secondary constraint $\cQ$ is preserved
in time. We again have
\begin{equation*}
\pb{\cQ(\bx),H}\approx \pb{\cQ(\bx),\HC{N}}
\end{equation*}
and thus the consistency condition for $\cQ$ requires that the Poisson
bracket between $\cQ$ and $\HC{N}$ must be a constraint (or zero).
No further constraints are required, since we obtain
\begin{equation}\label{pb:Q,HC}
\pb{\cQ(\bx),\HC{N}} = N\cH_0 (\bx) +ND^iD_i\cP(\bx) +3D_iN D^i\cP(\bx)
+2D^iD_iN\cP(\bx) \approx 0 \,.
\end{equation}
See \eqref{pb:Q,HC.Weyl+GR} in Appendix~\ref{appendix4} for the
derivation of this result, including all the Poisson brackets between
the Hamiltonian constraint and the canonical variables.

Since $\cP$ and $\cQ$ are first-class constraints, they generate
symmetry transformations.
We again introduce smeared versions of the constraints
\begin{equation}
\cP[\epsilon]=\int_{\Sigma_t}d^3x\epsilon\cP \,,\qquad
\cQ[\epsilon]=\int_{\Sigma_t}d^3x\epsilon\cQ \,,
\end{equation}
which are the generators.
The constraint $\cQ$ generates a scale transformation for the following
dynamical variables:
\begin{align}
\pb{h_{ij}(\bx),\cQ[\epsilon]}&=2\epsilon h_{ij}(\bx)\,,&
\pb{p^{ij}(\bx),\cQ[\epsilon]}&=-2\epsilon p^{ij}(\bx)\,,\\
\pb{K_{ij}(\bx),\cQ[\epsilon]}&=\epsilon K_{ij}(\bx)\,,&
\pb{\cP^{ij}(\bx),\cQ[\epsilon]}&=-\epsilon\cP^{ij}(\bx)
\,.\nn
\end{align}
Thus it is the generator of the conformal transformations.
We could easily extend $\cQ$ to a generator of scale transformations
for all variables, just as we did above for the momentum constraint, by
including the generators for the variables $N, N^i$ and their conjugated
momenta as $p_NN+p_iN^i$. Note that the conformal transformation leaves
the scalars $p$ and $\cP^{ij}K_{ij}$ invariant, which implies $\cQ$
itself is invariant, $\pb{\cQ(\bx),\cQ[\epsilon]}=0$. $\cP$ is simply
scaled under this transformation,
$\pb{\cP(\bx),\cQ[\epsilon]}=\epsilon\cP(\bx)$. On the other hand, $\cP$
generates a rather peculiar transformation
\begin{align}
\pb{h_{ij}(\bx),\cP[\epsilon]}&=0\,,&
\pb{p^{ij}(\bx),\cP[\epsilon]}&=-\epsilon
\cP^{ij}(\bx)\,,\\
\pb{K_{ij}(\bx),\cP[\epsilon]}&=\epsilon h_{ij}(\bx)\,,&
\pb{\cP^{ij}(\bx),\cP[\epsilon]}&=0\,.\nn
\end{align}
It evidently transforms $\cQ$ to $\cP$,
$\pb{\cQ(\bx),\cP[\epsilon]}=-\epsilon\cP(\bx)$.

The Hamiltonian constraint $\cH_0$ is preserved under time evolution,
since its Poisson bracket with itself is a sum of the momentum and
$\cP$ constraints (see Appendix~\ref{appendix4.1}):
\begin{equation}
\pb{\HC{\xi},\HC{\eta}}= \MC{\xi\vec{D}\eta-\eta\vec{D}\xi}
+2\cP\bigl[ \left( \xi D_i\eta-\eta D_i\xi \right) h^{ij}
\bigl( D^kK_{jk}-D_jK \bigr) \bigr] \,.
\end{equation}
This ensures that the time evolution of the system is consistent with
the structure of spacetime.

Since $\cQ$ is a first-class constraint, we should include it into the
total Hamiltonian with an arbitrary Lagrange multiplier:
\begin{equation}\label{H.Weyl.2}
H = \int_{\Sigma_t}d^3x\left( N\cH_0 +N^i\cH_i + \lambda_N p_N
+ \lambda^ip_i + \lambda_\cP\cP + \lambda_\cQ\cQ \right)
+\Hsurface \,.
\end{equation}

\subsubsection{Physical degrees of freedom and gauge fixing}
The number of physical degrees of freedom in any constrained system can
be counted according to Dirac's formula:
\begin{equation}\label{Dirac's formula}
\begin{split}
\text{Number of physical degrees of freedom} &= \frac{1}{2}(
\text{Number of canonical variables} \\
&- 2\times\text{Number of first-class constraints} \\
&- \text{Number of second-class constraints} )\,.
\end{split}
\end{equation}
In the Hamiltonian formulation of Weyl gravity, there are 32 canonical
variables, namely $N,N^i,h_{ij},K_{ij}$ and their canonically conjugated
momenta $p_N,p_i,p^{ij},\cP^{ij}$. There exist ten first-class
constraints, namely $p_N,p_i,\cH_0,\cH_i,\cP,\cQ$ and no second-class
constraints. Thus the conformally invariant Weyl gravity has six
physical degrees of freedom.

There exist many possible sets of gauge fixing conditions.
The simplest way to fix the gauge freedom associated with the primary
constraints $p_N=0$ and $p_i=0$, is to impose the lapse and shift
variables to constant values everywhere. There do exist useful
field-dependent choices for the conditions on $N$ and $N^i$, but we do
not consider them here.
Hence we impose the conditions
\begin{equation}\label{gauge.N}
\sigma_0=N-1=0\,,\qquad \sigma_i=N^i=0\,.
\end{equation}
The gauge freedom associated with the Hamiltonian and momentum
constraints $\cH_0=0$ and $\cH_i=0$ can be fixed by introducing four
conditions among the components of the metric $h_{ij}$. The gauge
freedom associated with the constraints $\cP=0$ and $\cQ=0$ can be fixed
by imposing the traces of the metric and the extrinsic curvature to
match those of the flat space.
Thus we can choose the gauge conditions as
\begin{align}
\chi_\mu(h_{ij})&=0\,,\qquad \mu=1,\ldots,4\,,\label{gauge.h}\\
K&=0\,,\qquad \chi_5=\delta^{ij}h_{ij}-3=0\,.\label{gauge.trace}
\end{align}
The four gauge conditions $\chi_\mu=0$, $\mu=1,\ldots,4$, have to be
such that they fix four components of the metric $h_{ij}$. These
conditions are often referred to as coordinate conditions. This is
because the conditions \eqref{gauge.N} and \eqref{gauge.h} essentially
fix the coordinate system on spacetime and define how the spacetime is
foliated.

An alternative choice of gauge fixing conditions, which is specific to
Weyl gravity, is to replace the five conditions $\chi_\mu(h_{ij})=0$,
$\mu=1,\ldots,5$, with conditions on the traceless component of the
extrinsic curvature $\bar{K}_{ij}$, i.e., we replace \eqref{gauge.h} and
\eqref{gauge.trace} with
\begin{equation}\label{gauge.K}
K=0\,,\qquad \chi_\mu(\bar{K}_{ij})=0\,,\qquad \mu=1,\ldots,5\,.
\end{equation}
The conditions $\chi_\mu(\bar{K}_{ij})=0$ have to be such that they fix
each of the five independent components of $\bar{K}_{ij}$.
This type of gauge is possible since the first-class constraints
depend on all the components of the variables $K_{ij},\cP^{ij}$, as
well as on all the components of the variables $h_{ij},p^{ij}$.
This enables a highly rich set of choices in the gauge fixing.
When a gauge of the type \eqref{gauge.K} is chosen, we may regard that
the 12 constraints define the variables $K_{ij},\cP^{ij}$
in terms of the independent variables $h_{ij},p^{ij}$.
For details on gauge fixing conditions, see \cite{York:1972,Smarr:1978i}
and the last reference in \cite{ADMreview}.

\subsection{Weyl gravity with \texorpdfstring{$\Lambda\neq0$}
{Lambda}}\label{sec4.2}
In this subsection, we consider what happens to the Hamiltonian
structure of Weyl gravity when the cosmological constant $\Lambda$ is
added into the Lagrangian.
The cosmological constant term is added into the potential part of the
Hamiltonian constraint as
\begin{equation}\label{H_0.Weyl+CC}
\begin{split}
\cH_0 &= 2p^{ij}K_{ij} -\frac{\cP_{ij}\cP^{ij}}{2\alpha\sqrt{h}}
+\cP^{ij}\sR_{ij} +\cP^{ij}K_{ij}K +D_iD_j\cP^{ij}\\
&\quad -\sqrt{h}\Lambda -\alpha\sqrt{h}
C_{ijk\bn}C^{ijk}_{\phantom{ijk}\bn} \,.
\end{split}
\end{equation}
All the primary constraints, the momentum constraint and the secondary 
constraint $\cQ$ remain the same as in the conformally invariant Weyl 
gravity.

The consistency condition that ensures the secondary constraint $\cQ$
to be preserved in time, now includes a cosmological constant term in
addition to the terms involving the constraints $\cH_0$ and $\cP$:
\begin{equation}\label{pb:cQ,H.wLambda}
\pb{\cQ(\bx),H}\approx \pb{\cQ(\bx),\HC{N}}
\approx 4\Lambda N\sqrt{h}(\bx) \,.
\end{equation}
Thus, whenever $\Lambda\neq0$, we have to introduce another secondary
constraint
\begin{equation}\label{cN}
\cN=N\sqrt{h} \approx 0 \,.
\end{equation}
In order to ensure the preservation of this constraint,
\begin{equation}
\pb{\cN(\bx),H}=NK\cN(\bx) +\lambda_N\sqrt{h} \approx \lambda_N\sqrt{h}
\,,
\end{equation}
we set the Lagrange multiplier of the primary constraint $p_N$ to zero,
\begin{equation}
\lambda_N=0 \,.
\end{equation}
Thus we have a pair of second-class constraints $\cN$ and $p_N$.
The lapse does not evolve, $\p_tN=0$, i.e., it is frozen to its initial
configuration. The Dirac bracket \eqref{Dirac-bracket.def} is equivalent
to the Poisson bracket for any quantities that are independent of $N$
and $p_N$. $\cP$ and $\cQ$ are still first-class constraints:
\begin{equation}
\pb{\cN(\bx),\cP(\by)}=0 \,,\qquad \pb{\cN(\bx),\cQ(\by)}
=3\cN(\bx)\delta(\bx-\by) \,.
\end{equation}
The number of physical degrees of freedom is six, similar to the pure
Weyl gravity. Unfortunately, the constraint \eqref{cN} is physically
unacceptable. The constraint \eqref{cN} imposes the determinant of the
metric of spacetime to be zero, $N\sqrt{h}=\sqrt{-g}=0$. This destroys
the geometry of spacetime.

For completeness, let us analyze the other possible secondary
constraints in place of \eqref{cN}.
If we impose the constraint $N\approx0$, then $N$ and $p_N$ become a
pair of second-class constraints. However, the time dimension collapses
when $N=0$. Recall that $N$ must be positive since $Ndt$ measures the
lapse of proper time between the hypersurfaces at times $t$ and $t+dt$.

Suppose we instead satisfy the condition \eqref{cN} by imposing the
constraint
\begin{equation}\label{Q_(2)}
\cQ_{(2)}=\sqrt{h} \approx 0 \,.
\end{equation}
This constraint has a weakly vanishing Poisson bracket with the
Hamiltonian constraint $\HC{N}$
\eqref{pb:h_ii,HC},
\begin{equation}
\pb{\cQ_{(2)}(\bx),\HC{N}}=N\sqrt{h}K(\bx)
=NK\cQ_{(2)}(\bx) \approx 0  \,.
\end{equation}
The Poisson brackets with $\cP$ and $\cQ$ are
\begin{align}
\pb{\cQ_{(2)}(\bx),\cP(\by)}&=0 \,,\\
\pb{\cQ_{(2)}(\bx),\cQ(\by)}&=
2h_{ij}(\by) \pb{\sqrt{h}(\bx),p^{ij}(\by)}
=3\cQ_{(2)}(\bx)\delta(\bx-\by) \,.\nn
\end{align}
Thus the Poisson bracket between the secondary constraint $\cQ_{(2)}$
and the Hamiltonian is proportional to $\cQ_{(2)}$, and hence no
further constraints are required. All the constraints appear to be
first class, which is very strange, since we would expect to see some
second-class constraint due to the violation of the conformal
invariance. The extra first-class constraint $\cQ_{(2)}$ implies the
removal of one physical degree of freedom. Certainly the introduction of
$\Lambda$ into Weyl gravity should not remove any physical degrees of
freedom! Compare this to Sec.~3.6 of \cite{Buchbinder:1987},
where four second-class constraints were found instead, denoted as
$C^{(k)}\approx0$, $k=1,\ldots,4$. It is however unclear why the
constraint $C^{(4)}$ is required in \cite{Buchbinder:1987}, since
$C^{(4)}$ is proportional to the constraint $C^{(3)}$ multiplied by $K$,
and thus $C^{(4)}$ is redundant. With this redundant 
fourth second-class constraint $C^{(4)}$, the same number of physical
degrees of freedom as in Weyl gravity were obtained in
\cite{Buchbinder:1987}, that is six physical degrees
of freedom. Our conclusion is the opposite: there exists one more
first-class constraint compared to the Weyl gravity case, $\cQ_{(2)}$,
and thus the number of physical degrees of freedom is five.

The constraint \eqref{Q_(2)} would generate a trivial (null)
transformation. The transformations of most of the variables are
zero strongly, while the only nontrivial transformation is that of
the momentum $p^{ij}$,
\begin{equation}
\pb{p^{ij}(\bx),\cQ_{(2)}[\epsilon]}=-\epsilon(\bx)\frac{\Lambda}{2}
\sqrt{h}h^{ij}(\bx)=-\frac{1}{2}h^{ij}\epsilon\cQ_{(2)}(\bx)
\approx0\,,
\end{equation}
and even that vanishes weakly.

The constraint \eqref{Q_(2)} would enable us to write the Hamiltonian
as
\begin{equation}\label{H.Weyl+CC}
H = \int_{\Sigma_t}d^3x\left( N\cH_0 +N^i\cH_i + \lambda_N p_N
+ \lambda^ip_i + \lambda_\cP\cP + \lambda_\cQ\cQ
+ \lambda_\cQ^{(2)}\cQ_{(2)} \right) +\Hsurface \,,
\end{equation}
where the Hamiltonian constraint can now be written as
\begin{equation}\label{H_0.Weyl+CC.2}
\cH_0= 2p^{ij}K_{ij} -\frac{\cP_{ij}\cP^{ij}}{2\alpha\sqrt{h}}
+\cP^{ij}\sR_{ij} +\cP^{ij}K_{ij}K +D_iD_j\cP^{ij}\,.
\end{equation}
As long as $\Lambda\neq0$, the terms that are proportional to $\sqrt{h}$
can be absorbed into the Lagrange multiplier term
$\lambda_\cQ^{(2)}\cQ_{(2)}$. The term
$-\frac{\cP_{ij}\cP^{ij}}{2\alpha\sqrt{h}}$, however, appears to be
divergent, since $h^{-1/2}\rightarrow\infty$.
It is not surprising that the constraint \eqref{Q_(2)} leads to such
inconsistencies, since the metric of the spatial hypersurfaces
$\Sigma_t$ must be positive definite, $h=\det(h_{ij})>0$.

Thus every one of the possible secondary constraints \eqref{cN},
$N\approx0$ or \eqref{Q_(2)} implies a physically inconsistent
Hamiltonian structure.

\subsection{Including \texorpdfstring{$\Lambda$}{Lambda}
into Weyl gravity with a scalar field}\label{sec4.3}
In order to resolve the previous problem with $\Lambda\neq0$, following
\cite{Antoniadis:1984kd,Antoniadis:1984br} we introduce a scalar field
$\phi$ which is coupled to the metric of spacetime in a way that makes
the theory invariant under conformal transformations, when the field
$\phi$ transforms in an appropriate way. This is in a sense reminiscent
of the introduction of gauge fields in order to ensure the invariance
under local phase transformations.
 Since the theory is required to possess both the conformal and
diffeomorphism invariances, the action for the scalar field must have
the form \cite{Antoniadis:1984kd,Antoniadis:1984br}
\begin{equation}\label{Sphi}
S_{\phi}=\int d^4x\sqrt{-g}\left[
\frac{1}{2}g^{\mu\nu}\partial_\mu\phi\partial_\nu\phi
+ \frac{1}{12}R\phi^2+\bar{\Lambda}\phi^4 \right] .
\end{equation}
We require that the scalar field transforms under conformal
transformations as $\phi\rightarrow \Omega^{-1}\phi$, while the metric
transforms in the usual way, $g_{\mu\nu}\rightarrow \Omega^2
g_{\mu\nu}$. The scalar curvature $R$ transforms as
\begin{equation}
R\rightarrow \Omega^{-2}
\left(R-6g^{\mu\nu}\frac{\nabla_\mu\nabla_\nu\Omega}{\Omega}\right).
\end{equation}
The value of $\bar{\Lambda}$ can be chosen freely, since it is not
fixed by conformal invariance. Hence the action \eqref{Sphi} is found to
be conformally invariant.
Consequently, the whole action of the theory,
$\SWeylLambda=\SWeyl+S_\phi$,
is conformally invariant as well.

In the action \eqref{Sphi}, notice that $\bar{\Lambda}$ is
dimensionless, while the cosmological constant $\Lambda$ has the
dimension $M^4$, mass to the fourth power.
The conformal invariance is broken spontaneously
\cite{Englert:1976ep,Englert:1975wj}, when the scalar field $\phi$ has a
nonzero vacuum expectation value. Naively, that would produce an
effective cosmological constant as $\Lambda=\bar{\Lambda}\bar{\phi}^4$,
where $\bar{\phi}$ is the vacuum expectation value of $\phi$. However,
it has been shown that the cosmological constant can be made to vanish
at every order in perturbation theory
\cite{Antoniadis:1984kd,Antoniadis:1984br}, even though
$\bar{\phi}\neq0$ is required for the existence of a perturbation
expansion.

For the Hamiltonian formulation we rewrite the action \eqref{Sphi} in
the $3+1$ form:
\begin{equation}
\begin{split}
S_{\phi}&=\int dt\int_{\Sigma_t} d^3\bx N\sqrt{h}\left[
-\frac{1}{2}(\nabla_n\phi)^2
+\frac{1}{2}h^{ij}\partial_i\phi \partial_j\phi \right. \\
&\quad -\frac{1}{3}K\phi\nabla_n\phi
-\frac{1}{6\sqrt{h}}\partial_i(\sqrt{h} h^{ij}\partial_j\phi^2) \\
&\quad +\left.\frac{1}{12}\left(K_{ij}K^{ij}-K^2+\sR\right)\phi^2
+\bar{\Lambda}\phi^4 \right]
+\Ssurface \,,
\end{split}
\end{equation}
where $ \nabla_n\phi= \frac{1}{N}(\partial_t\phi-N^i\partial_i\phi)$ and
$\Ssurface$ contains the boundary terms that appear due to integrations
by parts.
Then the momentum conjugate to $\phi$ has the form
\begin{equation}
p_\phi=-\sqrt{h}\nabla_n\phi-\frac{1}{3}\sqrt{h}K\phi \,.
\end{equation}
The contribution of the scalar field to the Hamiltonian is
\begin{equation}\label{Hphi}
H_\phi=\int_{\Sigma_t}d^3x(p_\phi\partial_t\phi-\cL_{\phi})
=\int_{\Sigma_t}d^3x\left( N\mH_0^{\phi}+N^i\mH_i^{\phi} \right)
+\Hsurface^\phi ,
\end{equation}
where
\begin{align}\label{cHphi}
\mH_0^{\phi} &= -\frac{1}{2\sqrt{h}} \left(p_\phi+\frac{1}{3}\sqrt{h}K
\phi \right)^2- \frac{1}{2}\sqrt{h} h^{ij}\partial_i\phi \partial_j\phi
+\frac{1}{6}\partial_i(\sqrt{h} h^{ij}\partial_j\phi^2) \\
&\quad -\frac{1}{12}\sqrt{h}\left(K_{ij}K^{ij}-K^2+\sR\right)\phi^2
- \bar{\Lambda}\sqrt{h}\phi^4 \,, \nonumber \\
\mH_i^{\phi}&=p_\phi\partial_i\phi \,,
\end{align}
and $\Hsurface^\phi$ contains the boundary terms. Since we wish to
obtain a boundary contribution only on the spatial boundary $\cB_t$,
we complement the action \eqref{Sphi} with a boundary term
$\frac{1}{6}\oint_{\p\cM}d^3x\sqrt{|\gamma|}K\phi^2$.
The variations of this boundary term are proportional to the variations
of
the variables and hence vanish due to the boundary conditions.
The boundary term in the Hamiltonian is obtained as
\begin{equation}
\Hsurface^\phi=-\frac{1}{6}\oint_{\cB_t}d^2x N\sqrt{\sigma}\left(
\Kt\phi^2 + s_ih^{ij}\p_j\phi^2 \right).
\end{equation}
The first term is similar to the boundary term of general relativity,
but weighted by the scalar field factor $\frac{\kappa}{6}\phi^2$.
The second term involves the gradient of the scalar factor $\phi^2$
along the unit normal to the spatial boundary.

Preservation of the primary constraint $\cP\approx 0$ leads to the
following form of the secondary constraint $\cQ\approx 0$:
\begin{equation}
\cQ=2p+\mathcal{P}^{ij}K_{ij}-p_{\phi}\phi \ .
\end{equation}
The constraint $\cQ$ is found to be the first-class constraint
associated with the conformal symmetry. Indeed, we obtain
\begin{equation}
\pb{\cQ(\bx),\mH_0^{\phi}(\by)}=\mH_0^{\phi}(\bx)\delta(\bx-\by) \,.
\end{equation}

Now we can fix the constraint $\cQ\approx 0$ by introducing the
gauge fixing condition. Instead of the gauge condition $\chi_5$ in
\eqref{gauge.trace}, we may impose
\begin{equation}\label{gauge.phi}
\chi_\phi= \phi(\bx)-\phi_0=0 \,,\qquad \phi_0=\mathrm{const} \,.
\end{equation}
Then $\cQ$ and $\chi_\phi$ become the second-class constraints that
vanish strongly and can be explicitly solved as
\begin{equation}
p_\phi=\frac{1}{\phi_0}\left(2p+\cP^{ij}K_{ij}\right).
\end{equation}
Note that the Dirac brackets between remaining phase space variables
are the same as the Poisson brackets, since they have vanishing Poisson
brackets with $\chi_\phi$.

The number of physical degrees of freedom is seven -- one more than in
the pure Weyl gravity. When the conformal gauge is fixed as in
\eqref{gauge.phi}, the extra scalar degree of freedom is transferred
to the metric variables. Alternatively, we can fix the gauge as in
\eqref{gauge.trace}, keeping the scalar variables $\phi,p_\phi$.

We emphasize that the kinetic term of $\phi$ in the action \eqref{Sphi}
has the opposite sign compared to a regular scalar field. As a result,
in the Hamiltonian \eqref{cHphi}, the kinetic term of $\phi$ is
nonpositive.

\subsection{General relativity plus Weyl gravity\texorpdfstring{:
$\kappa^{-1}\neq0$, $\alpha\neq0$, $\beta=\gamma=0$}{}}\label{sec4.4}
Here we consider the sum of \EH and Weyl actions.
This model is most relevant at long distances, where the \EH action
linear in curvature is expected to dominate the behaviour of the theory,
while contribution of the Weyl action is suppressed by the higher-order
derivatives.
The cosmological constant can be either included or excluded, since its
presence has no impact on the fundamental Hamiltonian structure of the
theory when the \EH action is included.

The Hamiltonian constraint is given as
\begin{equation}\label{H_0.Weyl+GR}
\begin{split}
\cH_0 &= 2p^{ij}K_{ij} -\frac{\cP_{ij}\cP^{ij}}{2\alpha\sqrt{h}}
+\cP^{ij}\sR_{ij} +\cP^{ij}K_{ij}K +D_iD_j\cP^{ij} \\
&\quad -\sqrt{h}\Lambda
-\frac{\sqrt{h}}{2\kappa}\left( \sR+K_{ij}K^{ij}-K^2 \right)
-\alpha\sqrt{h}C_{ijk\bn}C^{ijk}_{\phantom{ijk}\bn} \,.
\end{split}
\end{equation}
The surface term $-\frac{1}{\kappa}\oint_{\cB_t}d^2x
\sqrt{\sigma}\,N\Kt$ is now included in the total Hamiltonian
\eqref{H.Weyl.2} due to the presence of the \EH action  [see
\eqref{S_surface.EH.ort}]. In case the spacelike and timelike
hypersurfaces $\Sigma_t$ and $\cB$ intersect nonorthogonally, we would
include a surface term according to \eqref{S_surface.EH.nonort}.
Assuming the hypersurfaces are orthogonal, the surface term in the
Hamiltonian is written as
\begin{multline}\label{H_surf.Weyl+GR}
\Hsurface= -\frac{1}{\kappa}\oint_{\cB_t}d^2x\sqrt{\sigma}\,N\Kt
+\oint_{\cB_t}d^2x s_i\left( D_jN\cP^{ij}-ND_j\cP^{ij} \right.\\
+\left. 2N_jp^{ij}+2N^j\cP^{ik}K_{jk} \right).
\end{multline}

The secondary constraint $\cQ$ now takes a different form,
\begin{equation}\label{Q.Weyl+GR}
\cQ=2p+\cP^{ij}K_{ij}+\frac{2}{\kappa}\sqrt{h}K\approx0\,,
\end{equation}
because of the presence of the \EH part of the action.
The Poisson bracket between $\cP$ and $\cQ$ no longer closes,
\begin{equation}
\pb{\cP(\bx),\cQ(\by)}=\left( \cP -\frac{6}{\kappa}\sqrt{h} \right)(\by)
\delta(\bx-\by)\,.
\end{equation}
Clearly the conformal symmetry of Weyl gravity has been broken.
As a result, the consistency conditions that ensure the constraints
$\cP$ and $\cQ$ to be preserved in time, determine the Lagrange
multipliers of these constraints as
\begin{equation}
\lambda_\cP=-N\left[ \frac{2\kappa\Lambda}{3} +\frac{1}{2}
\left( \sR -K_{ij}K^{ij} +K^2 \right) \right]
\end{equation}
and
\begin{equation}
\lambda_\cQ=0\,.
\end{equation}
Thus $\cP$ and $\cQ$ are now second-class constraints.

Recall that in the Hamiltonian formalism, second-class constraints become strong
equalities if we replace the canonical Poisson bracket with the Dirac
bracket. Given a set of second-class constraints $\phi_a$,
$a=1,2,\ldots,A$, the Dirac bracket is defined as
\begin{multline}\label{Dirac-bracket.def}
\db{f_1(\bx),f_2(\by)}=\pb{f_1(\bx),f_2(\by)} -\iint_{\Sigma_t}d^3zd^3z'
\sum_{a,b=1}^A\pb{f_1(\bx),\phi_a(\bz)}M^{-1}_{ab}(\bz,\bz') \\
\times \pb{\phi_b(\bz'),f_2(\by)} ,
\end{multline}
where $M^{-1}(\bx,\by)$ is the inverse of the matrix $M(\bx,\by)$ with
the components
\begin{equation}\label{M.def}
M_{ab}(\bx,\by)=\pb{\phi_a(\bx),\phi_b(\by)},\qquad a,b=1,2,\ldots,A\,.
\end{equation}

The constraints $\cP$ and $\cQ$ can be set to zero strongly, when we
replace the Poisson bracket with the Dirac bracket.

The Dirac bracket between the canonical variables is defined as
\begin{align}
\db{h_{ij}(\bx),h_{kl}(\by)}&=0\,,\\
\db{h_{ij}(\bx),p^{kl}(\by)}&=\left( \delta_i^{(k}\delta_j^{l)}
+\frac{\kappa}{3}\frac{h_{ij}\cP^{kl}}{\sqrt{h}} \right) (\bx)
\delta(\bx-\by) \,,\nn\\
\db{h_{ij}(\bx),K_{kl}(\by)}&=-\frac{\kappa}{3}\frac{h_{ij}h_{kl}}
{\sqrt{h}} (\bx) \delta(\bx-\by) \,,\nn\\
\db{h_{ij}(\bx),\cP^{kl}(\by)}&=0\,,\nn\\
\db{p^{ij}(\bx),p^{kl}(\by)}&=\left[ \frac{\kappa}{3}
\frac{\cP^{ij}p^{kl} -p^{ij}\cP^{kl}}{\sqrt{h}}
+\frac{1}{6}K\left( \cP^{ij}h^{kl}-h^{ij}\cP^{kl} \right) \right]
 (\bx) \delta(\bx-\by) \,,\nn\\
\db{p^{ij}(\bx),K_{kl}(\by)}&=\left[ \frac{\kappa}{3}
\frac{p^{ij}h_{kl} -\frac{1}{2}\cP^{ij}K_{kl}}{\sqrt{h}}
+\frac{1}{6}h^{ij}h_{kl}K \right] (\bx) \delta(\bx-\by) \,,\nn\\
\db{p^{ij}(\bx),\cP^{kl}(\by)}&=\frac{\kappa}{6}
\frac{\cP^{ij}\cP^{kl}}{\sqrt{h}} (\bx) \delta(\bx-\by)  \,,\nn\\
\db{K_{ij}(\bx),K_{kl}(\by)}&=\frac{\kappa}{6}
\frac{h_{ij}K_{kl}-K_{ij}h_{kl}}{\sqrt{h}} (\bx) \delta(\bx-\by)
\,,\nn\\
\db{K_{ij}(\bx),\cP^{kl}(\by)}&=\left( \delta_i^{(k}\delta_j^{l)}
-\frac{1}{3}h_{ij}h^{kl} -\frac{\kappa}{6}
\frac{h_{ij}\cP^{kl}}{\sqrt{h}} \right) (\bx) \delta(\bx-\by) \,,\nn\\
\db{\cP^{ij}(\bx),\cP^{kl}(\by)}&=0\,.\nn
\end{align}
The total Hamiltonian is now written as
\begin{equation}\label{H.Weyl+GR}
H = \int_{\Sigma_t}d^3x\left( N\cH_0 +N^i\cH_i +\lambda_N p_N
+\lambda^ip_i \right) +\Hsurface \,.
\end{equation}

In the Hamiltonian formulation of the combination of Weyl and \EH
actions, there are 32 canonical variables ($N,N^i,h_{ij},K_{ij}$) and
their canonically conjugated momenta ($p_N,p_i,p^{ij},\cP^{ij}$). There
exist eight first-class constraints ($p_N,p_i,\cH_0,\cH_i$) and
two second-class constraints ($\cP,\cQ$). Thus the number of physical
degrees of freedom is seven. Gauge fixing can be accomplished similarly
as in Weyl gravity, but without the gauge conditions \eqref{gauge.trace}
which are associated with conformal invariance. For example, we can
choose the gauge conditions as in \eqref{gauge.N} and \eqref{gauge.h}.

We can now gain insight on the generality of the critical gravity
proposal \cite{Lu:2011zk}. In the full nonlinear theory, the value of
$\Lambda$ has no impact on the structure of the constraints and the
Hamiltonian. Since there exist eight first-class constraints
($p_N,p_i,\cH_0,\cH_i$) and two second-class constraints ($\cP,\cQ$),
regardless of the presence or value of $\Lambda$, the number of local
physical degrees of freedom is seven.
This suggests that the possibility for the massive spin-2 excitations
to become massless \cite{Lu:2011zk} is only an artefact of the
linearized theory on the anti-de Sitter spacetime.
This is likely related to the possibility of partial masslessness of
higher spin fields on (anti-)de Sitter backgrounds
\cite{Deser:1983mm,Deser:2001pe}.
In the linearized theory on Minkowski background, two modes are
associated with the massless spin-2 graviton and five modes with a
massive spin-2 field.

We have discovered a somewhat similar contrast between the linearized
formulation of the so-called renormalizable covariant gravity on
Minkowski spacetime and the Hamiltonian formulation of the full
nonlinear theory in \cite{Chaichian:2012md}.

\subsection{Curvature-squared gravity without conformal invariance}
\label{sec4.5}
Here we consider the gravitational action \eqref{S_C}, when the
curvature-squared part of the action lacks conformal invariance.
That is, we assume $\alpha\neq0$ and $\beta\neq0$ in the action
\eqref{S_C.dec}. Cosmological constant and \EH action can be either
included or excluded, since the conformal invariance is already broken
by the scalar curvature squared term.

The momentum \eqref{cP^ij} canonically conjugate to $K_{ij}$ can be
written as
\begin{equation}\label{cP^ij.R2}
\begin{split}
\cP^{ij}&=-\alpha\sqrt{h}\cG^{ijkl}\Lie{n}K_{kl}
+\alpha\sqrt{h}\bcG^{ijkl}\left( \sR_{kl} + K_{kl}K - \frac{1}{N}D_kD_lN
\right) \\
&\quad+\frac{\beta}{2}\sqrt{h}h^{ij}\left( \sR -3K_{kl}K^{kl} +K^2
-\frac{2}{N}D^kD_kN \right),
\end{split}
\end{equation}
where we have defined a generalized DeWitt metric as
\begin{equation}\label{cG^ijkl}
\cG^{ijkl}=\frac{1}{2}\left( h^{ik}h^{jl}+h^{il}h^{jk} \right)
-\frac{\alpha+3\beta}{3\alpha}h^{ij}h^{kl}\,.
\end{equation}
Since $\alpha\neq0$ and $\beta\neq0$, unlike the traceless DeWitt
metric \eqref{bcG^ijkl}, this generalized DeWitt metric
\eqref{cG^ijkl} has full rank, and hence its inverse can be obtained as
\begin{equation}\label{cG_ijkl}
\cG_{ijkl}=\frac{1}{2}\left( h_{ik}h_{jl}+h_{il}h_{jk} \right)
-\frac{\alpha+3\beta}{9\beta}h_{ij}h_{kl}\,.
\end{equation}
We can now solve the definition of the momentum \eqref{cP^ij.R2} in
terms of the velocities $\p_tK_{ij}$ and obtain
\begin{multline}
\cP^{ij}\p_tK_{ij}=-N\frac{\cP^{ij}\cG_{ijkl}\cP^{kl}}{\alpha\sqrt{h}}
+N\cP^{ij} \left( \sR_{ij} +K_{ij}K \right) +\cP^{ij}D_iD_jN\\
-\frac{N}{2}\cP \left( \sR -K_{ij}K^{ij} +K^2 \right)
+\cP^{ij}\cL_{\vec N}K_{ij} \,.
\end{multline}
The Lagrangian density of the action is written in terms of the
canonical variables as
\begin{multline}
\cL_C = -N\left( \frac{\cP^{ij}\cG_{ijkl}\cP^{kl}}{2\alpha\sqrt{h}}
+2p^{ij}K_{ij} -\sqrt{h}\Lambda -\frac{\sqrt{h}}{2\kappa}\left( \sR
+K_{ij}K^{ij} -K^2 \right) \right.\\
-\left.\alpha\sqrt{h}C_{ijk\bn}C^{ijk}_{\phantom{ijk}\bn} \right)
+p^{ij}\p_th_{ij} -2p^{ij}D_{(i}N_{j)}\,.
\end{multline}

The total Hamiltonian is obtained as
\begin{equation}\label{H.R2}
\begin{split}
H &= \int_{\Sigma_t}d^3x\left( N\cH_0 + N^i\cH_i + \lambda_N p_N
+ \lambda^ip_i \right) + \Hsurface \,,
\end{split}
\end{equation}
with the following quantities.
The Hamiltonian constraint is defined as
\begin{equation}\label{H_0.R2}
\begin{split}
\cH_0 &= 2p^{ij}K_{ij}
-\frac{\cP^{ij}\cG_{ijkl}\cP^{kl}}{2\alpha\sqrt{h}}
+\cP^{ij}\sR_{ij} +\cP^{ij}K_{ij}K +D_iD_j\cP^{ij}\\
&\quad -\frac{\cP}{2}\left( \sR -K_{ij}K^{ij} +K^2 \right)
 -\sqrt{h}\Lambda -\frac{\sqrt{h}}{2\kappa}\left( \sR +K_{ij}K^{ij} -K^2
\right) \\
&\quad -\alpha\sqrt{h}C_{ijk\bn}C^{ijk}_{\phantom{ijk}\bn} \,.
\end{split}
\end{equation}
The surface term is defined as
\begin{multline}
\Hsurface= -\frac{1}{\kappa}\oint_{\cB_t}d^2x\sqrt{\sigma}\,N\Kt
+\oint_{\cB_t}d^2x s_i\left( D_jN\cP^{ij}-ND_j\cP^{ij} \right.\\
+\left. 2N_jp^{ij}+2N^j\cP^{ik}K_{jk} \right),
\end{multline}
where the surface term of general relativity, i.e., the first term, is
included whenever the \EH action is included.

The algebra of constraints has the same form as in general relativity.
The Poisson bracket between Hamiltonian constraints is given as a sum of
momentum constraints with a $h^{ij}$-dependent multiplier,
\begin{equation}
\pb{\HC{\xi},\HC{\eta}}=\int_{\Sigma_t}d^3x\left( \xi D_i\eta-\eta
D_i\xi \right) h^{ij} \cH_j
=\MC{\xi\vec{D}\eta-\eta\vec{D}\xi} \,.
\end{equation}

In the Hamiltonian formulation of curvature-squared gravity without
conformal symmetry, and possibly with the cosmological constant and the
\EH action included, there are 32 canonical variables, namely
$N,N^i,h_{ij},K_{ij}$ and their canonically conjugated
momenta $p_N,p_i,p^{ij},\cP^{ij}$. There exist eight first-class
constraints, namely $p_N,p_i,\cH_0,\cH_i$, and no second-class
constraints. Thus the number of physical degrees of freedom is eight.

Gauge fixing can be accomplished in the same way as in the previous
case in Sec.~\ref{sec4.4}. For example, we can choose the gauge
conditions \eqref{gauge.N} and \eqref{gauge.h}.
Alternatively, we could impose the four gauge conditions \eqref{gauge.h}
on the variables $K_{ij}$ (or $\cP^{ij}$). But in these conformally
noninvariant theories, only part of the variables $K_{ij}$ can be
constrained, unlike in Weyl gravity \eqref{gauge.K}.

\subsection{Physical Hamiltonian and total gravitational energy}
For a system which is invariant under time reparameterization,
the Hamiltonian is typically a first-class constraint.
The same is true for generally covariant field theories with
diffeomorphism invariance. In a generally covariant system, time
evolution is just the unfolding of a gauge transformation.
The bulk part of the gravitational Hamiltonian is a sum of first-class
constraints, like in \eqref{H.Weyl} or in any other Hamiltonian
considered in this paper. However, the surface contribution $\Hsurface$
on the boundary of spatial hypersurface does not vanish on the
constraint surface. This indeed provides us the concept of total energy.

First, in order to obtain the physical Hamiltonian, we need to subtract
the reference background. Consider a given background solution and an
arbitrary (variable) configuration. The variable configuration and the
reference background should induce the same configuration on the spatial
boundary $\cB_t$, at least asymptotically. Hence the volume element on
the boundary is identical for them. Since the background is a solution
to the field equations, the constraints associated with the solution
vanish. Thus the Hamiltonian for the background consists solely of the
boundary terms $H_\mathrm{b}=H_\mathrm{b,\,surf}$.
The physical Hamiltonian is the difference
\begin{equation}
H_\mathrm{phys}=H-H_\mathrm{b}\,.
\end{equation}
Furthermore, for a stationary background solution, the canonical
momenta $p^{ij}_\mathrm{b}$ and $\cP^{ij}_\mathrm{b}$ vanish, since the
time derivatives of the variables $\p_th_{\mathrm{b},ij}$ and
$\p_tK_{\mathrm{b},ij}$ are zero. The spatial slices of the stationary
background can be labeled so that its lapse matches the lapse of the
variable configuration, $N_\mathrm{b}=N$.
Then the Hamiltonian of the background is obtained as follows:

\noindent (i) for pure Weyl or curvature-squared gravity:
\begin{equation}
H_\mathrm{b}=0\,,
\end{equation}
(ii) for Weyl gravity with $\Lambda$ included via a scalar field:
\begin{equation}
H_\mathrm{b}=-\frac{1}{6}\oint_{\cB_t}d^2x N\sqrt{\sigma}\left(
\Kt_\mathrm{b}\phi^2_\mathrm{b} +
s_ih^{ij}_\mathrm{b}\p_j\phi^2_\mathrm{b} \right),
\end{equation}
(iii) for general relativity with (or without) curvature-squared
terms:
\begin{equation}
H_\mathrm{b}=-\frac{1}{\kappa}\oint_{\cB_t}N\sqrt{\sigma}
\,\Kt_\mathrm{b}\,.
\end{equation}

We can now define the total energy associated with the time translation
along $t^\mu=Nn^\mu+N^\mu$  for any given solution of the equations
of motion as the value of the physical Hamiltonian on the constraint
surface. Since the constraint part of the Hamiltonian is zero for any
solution, the total energy is given by the difference of the surface
terms:
\begin{equation}\label{E}
E=\Hsurface-H_\mathrm{b} \,.
\end{equation}
We obtain:

\noindent (i) for pure Weyl or curvature-squared gravity:
\begin{equation}\label{E.R2}
E=\oint_{\cB_t}d^2x s_i\left( D_jN\cP^{ij}-ND_j\cP^{ij}
+2N_jp^{ij}+2N^j\cP^{ik}K_{jk} \right),
\end{equation}
(ii) for Weyl gravity with $\Lambda$ included via a scalar field:
\begin{equation}
\begin{split}\label{E.Weyl+phi}
E&=-\frac{1}{6}\oint_{\cB_t}d^2x N\sqrt{\sigma}\left[
\Kt\phi^2 -\Kt_\mathrm{b}\phi^2_\mathrm{b} + s_i\left( h^{ij}\p_j\phi^2
-h^{ij}_\mathrm{b}\p_j\phi^2_\mathrm{b} \right) \right]\\
&\quad +\oint_{\cB_t}d^2x s_i\left( D_jN\cP^{ij}-ND_j\cP^{ij}
+2N_jp^{ij}+2N^j\cP^{ik}K_{jk} \right),
\end{split}
\end{equation}
(iii) for general relativity with curvature-squared terms:
\begin{equation}\label{E.GR+R2}
\begin{split}
E&=-\frac{1}{\kappa}\oint_{\cB_t}d^2xN\sqrt{\sigma}\left( \Kt
-\Kt_\mathrm{b} \right)\\
&\quad +\oint_{\cB_t}d^2x s_i\left( D_jN\cP^{ij}-ND_j\cP^{ij}
+2N_jp^{ij}+2N^j\cP^{ik}K_{jk} \right).
\end{split}
\end{equation}
These generic expressions for the total energy can be used to obtain
the total energy with respect to different kinds of backgrounds,
as in general relativity \cite{Hawking:1995fd}. The two most relevant
cases being the asymptotically flat spacetimes \cite{Arnowitt:1962}
and the asymptotically anti-de Sitter spacetimes \cite{Abbott:1982npb}.
The energy formulae could also be generalized for nonorthogonally
intersecting boundaries $\Sigma_t$ and $\cB$, as in
\cite{Hawking:1996ww}.

The energy of pure curvature-squared gravity \eqref{E.R2} is
equivalent to the previous results in the literature.
When the \EH action is included, our expression for the total energy
\eqref{E.GR+R2} includes the familiar contribution of general
relativity. A physical interpretation is that the \EH term is expected
to dominate at great distances. In the case of Sec.~\ref{sec4.3}, the
energy \eqref{E.Weyl+phi} contains the contribution of the scalar field
$\phi$ thanks to its coupling to the scalar curvature of spacetime.

Recall that the total energy is always positive in general relativity
\cite{Schoen:1979zz,Witten:1981cmp}, except for flat Minkowski
spacetime, which has zero energy. Similarly, when the coupling
constants of the curvature-squared terms satisfy $\alpha\beta\le0$, the
total energy of curvature-squared gravity \eqref{E.R2} has been shown
to be zero for all exact solutions representing isolated systems
\cite{Boulware:1983prl} (see also \cite{Deser:2002rt}). This can be seen
as the result of energy confinement. The inclusion of the \EH term does
not change this feature. In fact the \EH contribution in \eqref{E.GR+R2}
is the dominate one, since the curvature falls off quicker in the
asymptotic region when the \EH term is included. In the case of
\eqref{E.Weyl+phi}, the asymptotic boundary condition for the scalar
field can be chosen so that the total energy resembles the case of
\eqref{E.GR+R2}: in an asymptotically flat spacetime, $\phi=C+O(r^{-b})$
where $C$ is a constant and $b>1$, so that the gradient term of $\phi$
is suppressed and $C^2/3$ takes the role of the gravitational constant
$\kappa^{-1}$.

\subsection{Alternative Hamiltonian formulations}
We emphasize that the first-order ADM forms of the actions and the
discussion of the boundary surface terms presented in Sec.~\ref{sec3},
as well as the following Hamiltonian analysis presented in
Sec.~\ref{sec4}, are specific to the chosen independent variables of the
action \eqref{S.second-order}. For any higher-derivative theory,
there exists many possible choices for the independent variables.
Since the higher-order derivatives imply the existence of extra degrees
of freedom, one introduces extra independent variables which carry the
extra degrees of freedom of the theory. The choice of independent
variables defines the form of the first-order action, which in turn
defines the Hamiltonian structure of the theory.
The different Hamiltonian formulations of a given higher-derivative
theory should be related by canonical transformations
\cite{Buchbinder:1987}. Hence they should be physically equivalent (at
least classically). For an example of an alternative Hamiltonian
formulation of higher-derivative gravity, see \cite{Deruelle:2009zk}.

Furthermore, the choice of boundary conditions is not unique.
For instance, if the curvature tensor of spacetime would be considered
as an independent variable, it would be natural to impose boundary
conditions on the curvature tensor. Since such variables and
their boundary conditions involve second-order derivatives, the
formulation would clearly differ from the present formulation,
where the extrinsic curvature \eqref{K_ij} is chosen to be
an independent variable of the first-order action.

\section{Conclusions}
We have presented Hamiltonian analysis of Weyl gravity and of other
fully diffeomorphism-invariant curvature-squared gravitational models.
We concentrated on the potentially renormalizable theories, whose
linearized actions are known to include notorious ghost fields with
negative energy. All the surface terms on the boundary of spacetime were
accounted for in each theory, as well as the freedom to include surface
terms that vanish due to the expanded configuration space of
higher-derivative gravity, which includes both the fundamental forms of
the hypersurfaces. The expression for the total energy was obtain in
each case with respect to a generic stationary background.

Compared to the seminal work in \cite{Boulware:1984}, a correction to
the component of Weyl tensor that is fully tangent to the spatial
hypersurface was discovered in \eqref{pC.ADM}. A fully traceless
component appears, namely the properly symmetrized, traceless, quadratic
extrinsic curvature tensor $\cK_{ijkl}$ defined in \eqref{cK.ADM}. The
square of $\cK_{ijkl}$ vanishes due to the Cayley-Hamilton theorem.
Hence the correction makes a difference in theories where the Weyl
tensor is coupled to something else than itself. But it does not appear
in the action of curvature-square gravity \eqref{S_C}.
Therefore the Hamiltonian structures presented in
Sections~\ref{sec4.1}, \ref{sec4.4}, \ref{sec4.5} are similar to those
found in the literature (see \cite{Boulware:1984,Buchbinder:1987,
Demaret:1995,Querella:1998,Deruelle:2009zk}).
The only relevant difference is the presence of the surface Hamiltonian
of general relativity -- the first term in \eqref{H_surf.Weyl+GR} --
when the \EH action is included ($\kappa^{-1}\neq0$). In that case,
the expression for total energy \eqref{E.GR+R2} is complemented by
the energy term of general relativity -- the first term in
\eqref{E.GR+R2} -- which is the dominant contribution in asymptotically
flat spacetimes.

We found in Sec.~\ref{sec4.2} that including a nonvanishing
cosmological constant into Weyl gravity implies a severe problem.
Since the determinant of the metric of spacetime is forced to zero by a
secondary constraint \eqref{cN}, the Hamiltonian structure becomes
physically inconsistent. Thus Weyl gravity with $\Lambda\neq0$ is not a
well-defined theory.

In Sec.~\ref{sec4.3}, we analyzed the possibility to include a scalar
field which is coupled to the scalar curvature of spacetime in way that
preserves conformal invariance
\cite{Antoniadis:1984kd,Antoniadis:1984br}. Conformal invariance is
broken spontaneously if the scalar field has a nonzero vacuum
expectation value, producing an effective \EH term and possibly a
cosmological constant. The kinetic term of the scalar field is
nonpositive, what may jeopardize the stability of the system.

In all the cases that include the Weyl action, i.e., when $\alpha\neq0$
in the action \eqref{S_C}, the  Ostrogradskian form of the Hamiltonian
is clearly visible in the first term $2p^{ij}K_{ij}$ of the Hamiltonian
constraint, which is linear in the momentum $p^{ij}$. This implies the
appearance of the Ostrogradskian instability. In the absence of
conformal invariance, there exist five or six unstable degrees of
freedom depending on whether $\beta=0$ or $\beta\neq0$, respectively,
which are associated with the five or six independent time derivatives
of the components of the extrinsic curvature on the spatial
hypersurface. Since there exists only four first-class constraints --
associated with the diffeomorphism invariance -- the constraints cannot
restrain the higher-derivative degrees of freedom.
Only in the case of conformally invariant Weyl gravity, there exist as
many constraints as there are unstable directions in phase space.
This follows from the fact that the Weyl action contains the
five independent traceless components of the time derivative of the
extrinsic curvature, and it possesses five first-class constraints
which are associated with the diffeomorphism and conformal invariance.
Hence in principle, only the conformally invariant Weyl gravity has
enough local constraints to be able to restrain the unstable degrees of
freedom. In all the other potentially renormalizable cases, the number
of independent second-order derivatives in the Lagrangian exceeds the
number of local invariances.
Thus Weyl gravity is the only potentially renormalizable theory of the
type \eqref{S_C} that might avoid the problem with instability,
which manifests itself as ghosts and lacks unitarity in the linearized
theory.\footnote{Recently \cite{Mannheim:2012}, it has been argued that
conformal gravity is unitary, but its Hamiltonian is non-Hermitian.
However, in order to achieve this, the gravitational field $g_{\mu\nu}$
would have to be anti-Hermitian, i.e., the metric would be purely
imaginary.}

However, perturbative analyses suggest that even Weyl gravity cannot
escape the ghost problem. On the flat background, linearized Weyl
gravity includes a massless spin-2 ghost \cite{Riegert:1984}. The
inclusion of \EH action implies the appearance of a massive spin-2
ghost, as well as a massive scalar ghost if $\beta\neq0$ in \eqref{S_C}.
The dilemma of generally covariant higher-derivative gravity is that the
spin-2 ghost is required for renormalizability \cite{Stelle:1977}.
In the full nonlinear theory, further study of the problem is still
required.

The recent claim of obtaining a critical case of curvature-squared
gravity \cite{Lu:2011zk}, where the spin-2 ghost becomes massless,
was concluded in our Sec.~\eqref{sec4.4} to be a specific feature of
the linearized theory on the anti-de Sitter background. In the full
nonlinear theory, however, it was shown in Sec.~\eqref{sec4.4} that
the number and nature of local physical degrees of freedom are
independent of the value of the cosmological constant, when both the \EH
and Weyl actions are included.

\paragraph{Acknowledgements}
We are indebted to Masud Chaichian for his interest in this work and for
many useful and clarifying discussions.
We thank Andrei Smilga for a fruitful communication.
The support of the Academy of Finland under the Projects
No. 136539 and No. 272919, as well as of the Magnus Ehrnrooth
Foundation, is gratefully acknowledged.
The work of J.K. was supported by the Grant Agency of the Czech
Republic under the grant P201/12/G028.
The work of M.O. was also supported by the Jenny and Antti Wihuri
Foundation.

\appendix

\section{Appendix: Notation}\label{appendix1}
The metric tensor $g_{\mu\nu}$ of spacetime has the signature
$(-,+,+,+)$.

Symmetrization and antisymmetrization of tensor indices is denoted by
parentheses and brackets, respectively.
Normalization is chosen so that the (anti)symmetrization has no effect
on an already (anti)symmetric tensor. For example, we denote
\begin{equation}
A_{(\mu\nu)} = \frac{1}{2}\left( A_{\mu\nu}+A_{\nu\mu}
\right),\qquad
A_{[\mu|\rho}B_{|\nu]}^{\phantom{|\nu)}\rho}
=\frac{1}{2}\left(
A_{\mu\rho}B_{\nu}^{\phantom{\nu}\rho}
-A_{\nu\rho}B_{\mu}^{\phantom{\mu}\rho} \right).
\end{equation}
We may also use the following notation
\begin{equation}\label{symmetrization2}
A_{\mu\nu}+(\mu\leftrightarrow\nu) = A_{\mu\nu}+A_{\nu\mu}
\,,\qquad A_{\mu\nu}-(\mu\leftrightarrow\nu) =
A_{\mu\nu}-A_{\nu\mu} \,,
\end{equation}
if it is more convenient than the one with parentheses and brackets.
This can be the case when $A_{\mu\nu}$ in \eqref{symmetrization2} is a
long expression containing several terms.
No normalization is included in this notation.
This notation may also be used to denote (anti)symmetrization with
respect to the exchange of functions.

\section{Appendix: The Cayley-Hamilton theorem}\label{appendix2}
The Cayley-Hamilton theorem states that any square matrix $A$ over a
commutative ring is the root of its own characteristic polynomial,
$P(A)=0$. The characteristic polynomial is defined as
$P(\lambda)=\det(\lambda I-A)$, where $I$ is the unit matrix.

The Cayley-Hamilton theorem has a tensor form due to the well known
relationship between matrices and linear transformations and rank 2
tensors on a vector space. Considering a tensor
$A^\mu_{\phantom\mu\nu}$ on a $d$-dimensional vector space, such as the
tangent space of a $d$-dimensional smooth manifold, the Cayley-Hamilton
theorem can be written as
\begin{equation}
\begin{split}
P(A)^\mu_{\phantom\mu\nu}&=-(d+1)\delta^\mu_{\phantom\mu[\nu}
A^{\rho_1}_{\phantom{\rho_1}\rho_1}
A^{\rho_2}_{\phantom{\rho_1}\rho_2}\cdots
A^{\rho_d}_{\phantom{\rho_1}\rho_d]} \\
&=\left( A^d \right)^\mu_{\phantom\mu\nu}
+c_1\left( A^{d-1} \right)^\mu_{\phantom\mu\nu}
+\ldots+c_{d-1}A^\mu_{\phantom\mu\nu} +c_d\delta^\mu_{\phantom\mu\nu}
=0\,,
\end{split}
\end{equation}
where the coefficients $c_n$ are given as
\begin{equation}
c_n=(-1)^n A^{\mu_1}_{\phantom{\mu_1}[\mu_1}
A^{\mu_2}_{\phantom{\mu_2}\mu_2}\cdots
A^{\mu_n}_{\phantom{\mu_n}\mu_n]}
\,,\qquad n=1,2,\ldots,d
\end{equation}
and we denote the tensor $A^\mu_{\phantom\mu\nu}$ to the $m$-th
power as
\begin{equation}
\left( A^m \right)^\mu_{\phantom\mu\nu}=
A^\mu_{\phantom\mu\rho_1}A^{\rho_1}_{\phantom{\rho_1}\rho_2}
\cdots A^{\rho_{m-2}}_{\phantom{\rho_{m-2}}\rho_{m-1}}
A^{\rho_{m-1}}_{\phantom{\rho_{m-1}}\nu}
\,,\qquad m=2,3,\ldots,d \,.
\end{equation}

We shall apply the Cayley-Hamilton theorem to a tensor field on the
three-dimensional Riemannian manifold $\Sigma_t$.
A tensor $A^i_{\phantom ij}$ on a three-dimensional vector space
satisfies
\begin{equation}\label{P(A)=0.3d}
\begin{split}
P(A)^i_{\phantom ij}&=
A^i_{\phantom ik}A^k_{\phantom kl}A^l_{\phantom lj}
-A^i_{\phantom ik}A^k_{\phantom kj}A
-\frac{1}{2}A^i_{\phantom ij}\left( A^k_{\phantom kl}A^l_{\phantom lk}
-A^2 \right) \\
&\quad -\frac{\delta^i_{\phantom ij}}{6} \left(
2A^k_{\phantom kl}A^l_{\phantom lm}A^m_{\phantom mk}
-3A^k_{\phantom kl}A^l_{\phantom lk}A +A^3 \right) =0 \,,
\end{split}
\end{equation}
where $A=A^i_{\phantom ii}$ denotes the trace.

\section{Appendix: Removing the auxiliary variables}\label{appendix3}
We show that the Hamiltonian formalism where the Lagrange multiplier
$\lambda^{ij}$ and its conjugated momentum $p^\lambda_{ij}$ are included
as canonical variables is equivalent to the formalism presented in
Sec.~\ref{sec4}. If the canonical variables $\lambda^{ij}$ and
$p^\lambda_{ij}$ are included, we obtain the extra primary constraints
\begin{equation}\label{pc.lambda}
\Pi^{ij}=p^{ij}-\sqrt{h}\lambda^{ij}\approx0\,,\qquad
p^\lambda_{ij}\approx0\,.
\end{equation}
Each of these constraints has a nonvanishing Poisson bracket
with one other constraint,
\begin{equation}\label{pb:Pi,p^lambda_ij}
\pb{\Pi^{ij}(\bx),p^\lambda_{kl}(\by)}=-\delta^{(i}_k \delta^{j)}_l
\sqrt{h}\delta(\bx-\by) \,.
\end{equation}
Thus $\Pi^{ij}$ and $p^\lambda_{ij}$ are second-class constraints.
Second-class constraints become strong equalities when we
replace the canonical Poisson bracket with the Dirac bracket
\eqref{Dirac-bracket.def}.
The matrix \eqref{M.def} for the second-class constraints
$\phi_a=(\Pi^{ij},p^\lambda_{ij})$ is given by the Poisson brackets
\eqref{pb:Pi,p^lambda_ij} in the cross-diagonal form
$M(\bx,\by)=\bigl(\begin{smallmatrix}0&-1\\ 1&0\end{smallmatrix}\bigr)
\sqrt{h}\delta(\bx-\by)$, where $0$ and $1$ denote the nine-dimensional
zero and unit matrices, respectively. The inverse matrix has the form
$M^{-1}(\bx,\by)=\bigl(\begin{smallmatrix}0&1\\
-1&0\end{smallmatrix}\bigr) \frac{1}{\sqrt{h}}\delta(\bx-\by)$
and the Dirac bracket is defined as
\begin{equation}\label{Dirac-bracket}
\begin{split}
\db{f_1(\bx),f_2(\by)}&=\pb{f_1(\bx),f_2(\by)} -\int_{\Sigma_t}d^3z
\frac{1}{\sqrt{h}}\pb{f_1(\bx),\Pi^{ij}(\bz)}\pb{p^\lambda_{ij}(\bz),
f_2(\by)} \\
&\qquad +\int_{\Sigma_t}d^3z \frac{1}{\sqrt{h}}
\pb{f_1(\bx),p^\lambda_{ij}(\bz)}\pb{\Pi^{ij}(\bz),f_2(\by)} .
\end{split}
\end{equation}
Then we set the constraints \eqref{pc.lambda} to zero strongly and
eliminate the variables $\lambda^{ij}$ and $p^\lambda_{ij}$ by
substituting
\begin{equation}\label{lambda=p}
\lambda^{ij}=\frac{p^{ij}}{\sqrt{h}} \,,\qquad p^\lambda_{ij}=0 \,.
\end{equation}
The Dirac bracket \eqref{Dirac-bracket} modifies the Poisson bracket if
one of the arguments depends on $\lambda^{ij}$ and the other argument
depends on $h_{ij}$, $p^{ij}$ or $p^\lambda_{ij}$. Otherwise the Dirac
bracket is equivalent to the Poisson bracket. Since we have solved the
constraints everywhere as \eqref{lambda=p}, none of the arguments can
depend on $\lambda^{ij}$ or $p^\lambda_{ij}$. Thus the Dirac bracket
\eqref{Dirac-bracket} is equivalent to the Poisson bracket
\begin{equation}\label{Db=Pb}
\db{f_1(\bx),f_2(\by)}=\pb{f_1(\bx),f_2(\by)} ,
\end{equation}
for any arguments $f_1$ and $f_2$ that depend on the remaining variables
$N,N^i,h_{ij},K_{ij}$ and their canonically conjugated momenta
$p_N,p_i,p^{ij},\cP^{ij}$. Therefore, introducing the Dirac bracket and
imposing the second-class constraints strongly is equivalent to
substituting \eqref{lambda=p} and removing the auxiliary
variables $\lambda^{ij}$ and $p^\lambda_{ij}$ from the system.

We can now see that it is unnecessary to include the Lagrange multiplier
$\lambda^{ij}$ and its conjugated momentum as extra canonical variables.
We can directly identify $\sqrt{h}\lambda^{ij}$ as the canonical
momentum $p^{ij}$ conjugate to $h_{ij}$ and hence avoid the inclusion
of extra canonical variables. This is a general feature of the
Hamiltonian formulation of higher-derivative theories (see e.g.
\cite{Buchbinder:1987}).

\section{Appendix: Calculation of Poisson brackets for the Hamiltonian
constraint\texorpdfstring{ $\cH_0$}{}}\label{appendix4}
Because of the simple $p^{ij}$ dependence of the Hamiltonian constraint
$\cH_0$, namely the term $2p^{ij}K_{ij}$, we have the following
Poisson brackets between the metric and $\HC{\xi}$:
\begin{align}
\pb{h_{ij}(\bx),\HC{\xi}}&=2\xi(\bx)K_{ij}(\bx) \,,\label{pb:h_ii,HC}\\
\pb{h^{ij}(\bx),\HC{\xi}}&=-2\xi(\bx)K^{ij}(\bx) \,,\nn\\
\pb{\sqrt{h}(\bx),\HC{\xi}}&=\xi(\bx)\sqrt{h}K(\bx)\,,\nn\\
\pb{\frac{1}{\sqrt{h}(\bx)},\HC{\xi}}&=-\xi(\bx)\frac{K}{\sqrt{h}}(\bx)
 \,.\nn
\end{align}
The rest of the Poisson brackets differ depending on which of the
couplings are switched on. In particular, including the $R^2$ term into
the Lagrangian \eqref{S_C}, i.e., $\beta\neq0$, alters the kinetic part
of the Hamiltonian significantly.

\subsection{Weyl gravity\texorpdfstring{: $\alpha\neq0$,
$\beta=\gamma=0$ in \eqref{S_C}}{}}\label{appendix4.1}
First we consider the cases where the curvature-squared part of the
action \eqref{S_C} is the Weyl action ($\alpha\neq0$, $\beta=\gamma=0$).
The \EH action and the cosmological constant can be included or
excluded, since they do not alter the kinetic part of the Hamiltonian
constraint. The Hamiltonian constraint is given in \eqref{H_0.Weyl+GR}.
When the cosmological constant and/or the \EH action are not present in
the action, one simply sets $\Lambda=0$ and/or $\kappa^{-1}=0$ in the
following results.

\subsubsection{Poisson brackets between the canonical variables and the
Hamiltonian constraint}\label{appendix4.1.1}
The Poisson bracket between $K_{ij}$ and the Hamiltonian constraint
$\HC{\xi}$ reads as
\begin{equation}\label{pb:K_ii,HC}
\pb{K_{ij}(\bx),\HC{\xi}}= \xi(\bx)\left(
-\frac{\cP_{ij}}{\alpha\sqrt{h}}
+\sR_{ij} +K_{ij}K \right)(\bx)
+D_{ij} \xi(\bx)\,.
\end{equation}
We shall denote the symmetrized second-order covariant derivative as
$D_{ij}=D_{(i}D_{j)}$ and later the symmetrized higher-order
covariant derivatives similarly as $D_{ijk}=D_{(i}D_jD_{k)}$ etc.

The Poisson bracket between the momentum $\cP^{ij}$ and the Hamiltonian
constraint $\HC{\xi}$ is obtained as
\begin{equation}\label{pb:HC,cP^ij}
\begin{split}
\pb{\HC{\xi},\cP^{ij}(\bx)} &= \xi(\bx) \left[ 2p^{ij}
+\cP^{ij}K +h^{ij}\cP^{kl}K_{kl}
-\frac{\sqrt{h}}{\kappa}\left( K^{ij}-h^{ij}K \right) \right] \\
&\quad + 4\alpha\sqrt{h}D_k\left( \xi C^{k(ij)}_{\phantom{k(ij)}\bn}
\right)(\bx) \,.
\end{split}
\end{equation}

The Poisson bracket between the momentum $p^{ij}$ and the Hamiltonian
constraint $\HC{\xi}$ is very complicated. It can be obtained after
a quite laborious calculation as
\begin{equation}\label{pb:HC,p^ij}
\pb{\HC{\xi},p^{ij}(\bx)}= E_{(0)}^{ij} \xi(\bx) +
E_{(1)}^{ijk}D_k\xi(\bx) +E_{(2)}^{ijkl}D_{kl}\xi(\bx) \,,
\end{equation}
where we have defined the three coefficient tensor densities
$E_{(I)}^{i_1\cdots i_{2+I}}$ ($I=0,1,2$) as
\begin{equation}
\begin{split}
E_{(0)}^{ij}&=-\frac{1}{\alpha\sqrt{h}}\left(
\cP^i_{\phantom{i}k}\cP^{jk}
 -\frac{1}{4}h^{ij}\cP_{kl}\cP^{kl} \right)
+D_kD^{(i}\cP^{j)k} \\
&\quad -\frac{1}{2}h^{ij}D_{kl}\cP^{kl}
-\frac{1}{2}D^kD_k\cP^{ij} -\cP^{kl}K_{kl}K^{ij}
-\frac{1}{2}\sqrt{h}h^{ij}\Lambda \\
&\quad +\frac{\sqrt{h}}{2\kappa}\left[ \sR^{ij}
+2K^i_{\phantom ik}K^{jk}-2K^{ij}K
-\frac{1}{2}h^{ij}\left( \sR+K_{ij}K^{ij}-K^2 \right) \right] \\
&\quad +\alpha\sqrt{h} \left[
2C^i_{\phantom{i}kl\bn}C^{jkl}_{\phantom{jkl}\bn}
+C_{kl\phantom{i}\bn}^{\phantom{kl}i}
C^{klj}_{\phantom{klj}\bn}
-\frac{1}{2}h^{ij}C_{klm\bn}C^{klm}_{\phantom{klm}\bn}
 \right.\\
&\quad -2D_kK_l^{\phantom l(i}C^{j)lk}_{\phantom{j)lk}\bn}
-2D_kK_l^{\phantom l(i|}C^{kl|j)}_{\phantom{kl|j)}\bn}
-2K_l^{\phantom l(i}D_kC^{j)lk}_{\phantom{j)lk}\bn}\\
&\quad -\left.2K_l^{\phantom l(i|}D_kC^{kl|j)}_{\phantom{kl|j)}\bn}
-2K_{kl}D^k C^{l(ij)}_{\phantom{l(ij)}\bn}
-2\left( 2D_lK^l_{\phantom lk} -D_kK \right)
C^{k(ij)}_{\phantom{k(ij)}\bn} \right],
\end{split}
\end{equation}
\begin{equation}
\begin{split}
E_{(1)}^{ijk}&=D^{(i}\cP^{j)k} +2h^{k(i}D_l\cP^{j)l}
-h^{ij}D_l\cP^{kl} -\frac{3}{2}D^k\cP^{ij} \\
&\quad -2\alpha\sqrt{h}\left( K_l^{\phantom l(i}
C^{j)lk}_{\phantom{j)lk}\bn}
+K_l^{\phantom l(i|}C^{kl|j)}_{\phantom{kl|j)}\bn}
+K^k_{\phantom kl}C^{l(ij)}_{\phantom{l(ij)}\bn} \right)
\end{split}
\end{equation}
and
\begin{equation}
E_{(2)}^{ijkl}=h^{i(k}\cP^{l)j} +h^{j(k}\cP^{l)i}
-\frac{1}{2}h^{ij}\cP^{kl} -h^{kl}\cP^{ij}
+\frac{\sqrt{h}}{2\kappa}\left( h^{ij}h^{kl}
-h^{i(k}h^{l)j} \right)  \,.
\end{equation}

In Weyl gravity, the Poisson bracket between $\cQ$ and $\HC{\xi}$
requires the trace of the Poisson bracket between $p^{ij}$ and
$\HC{\xi}$. It can be obtained from \eqref{pb:HC,p^ij} as
\begin{equation}\label{trace:pb:HC,p^ij}
\begin{split}
\pb{\HC{\xi},p^{ij}(\bx)}h_{ij}(\bx) &= -\xi(\bx) \left[
\frac{\cP_{ij}\cP^{ij}}{4\alpha\sqrt{h}}
 +\frac{1}{2}D_iD_j\cP^{ij} +\frac{1}{2}D^iD_i\cP
+\cP^{ij}K_{ij}K \right.\\
&\quad +\frac{3}{2}\sqrt{h}\Lambda +\frac{\sqrt{h}}{4\kappa}
\left( \sR -K_{ij}K^{ij} +K^2 \right) \\
&\quad +\left.\alpha\sqrt{h} \left(
2D_i C^{ijk}_{\phantom{ijk)}\bn}K_{jk}
-\frac{1}{2}C_{ijk\bn}C^{ijk}_{\phantom{ijk}\bn} \right)
\right] (\bx) \\
&\quad -D_i\xi(\bx) \left( \frac{3}{2}D^i\cP
+2\alpha\sqrt{h}C^{ijk}_{\phantom{ijk}\bn}K_{jk} \right)(\bx) \\
&\quad +D_{ij}\xi(\bx) \left( \frac{1}{2}\cP^{ij} -h^{ij}\cP
+\frac{\sqrt{h}h^{ij}}{\kappa} \right)(\bx) \,.
\end{split}
\end{equation}
The Poisson bracket between $\cQ$ and $\HC{\xi}$ will be obtained in the
next subsection, \ref{appendix4.1.2}.

\subsubsection{Poisson brackets between the Hamiltonian constraints and
with the other constraints}\label{appendix4.1.2}
Poisson brackets between the Hamiltonian constraint and the other
constraints are then determined by using the previous results for the
canonical variables . The Poisson bracket between the constraints $\cP$
and $\HC{\xi}$ is a sum of the constraints $\cP$ and $\cQ$:
\begin{equation}
\begin{split}
\pb{\cP(\bx),\HC{\xi}} &= \cP^{ij}(\bx)\pb{h_{ij}(\bx),\HC{\xi}}
-h_{ij}(\bx)\pb{\HC{\xi},\cP^{ij}(\bx)} \\
&=-\xi(\bx)\left( \cQ +\cP K \right)(\bx) \,,
\end{split}
\end{equation}
where $\cQ$ is defined by \eqref{Q.Weyl+GR}.
The Poisson bracket between the constraints $\cQ$ and $\HC{\xi}$ is
a sum of constraints only in the case of pure Weyl gravity. If the
cosmological constant and/or \EH term are included into the action, the
Poisson bracket between $\cQ$ and $\HC{\xi}$ is not a sum of
constraints. We then obtain the Poisson bracket as
\begin{equation}\label{pb:Q,HC.Weyl+GR}
\begin{split}
\pb{\cQ(\bx),\HC{\xi}} &= 2p^{ij}(\bx)\pb{h_{ij}(\bx),\HC{\xi}}
-2\pb{\HC{\xi},p^{ij}(\bx)}h_{ij}(\bx)\\
&\quad +\cP^{ij}(\bx)\pb{K_{ij}(\bx),\HC{\xi}}
-\pb{\HC{\xi},\cP^{ij}(\bx)}K_{ij}(\bx)\\
&\quad +\frac{2}{\kappa}\left( \pb{\sqrt{h}(\bx),\HC{\xi}}K(\bx)
+\sqrt{h}K_{ij}(\bx)\pb{h^{ij}(\bx),\HC{\xi}} \right. \\
&\quad +\left.\sqrt{h}h^{ij}(\bx)\pb{K_{ij}(\bx),\HC{\xi}} \right) \\
&=\xi\cH_0  +\xi D^iD_i\cP +3D_i\xi D^i\cP
+2D^iD_i\xi\cP -\xi\frac{2}{\kappa\alpha}\cP \\
&\quad +\xi\sqrt{h}\left[ 4\Lambda +\frac{3}{\kappa}
\left( \sR -K_{ij}K^{ij} +K^2 \right) \right] ,
\end{split}
\end{equation}
where we have omitted the arguments $(\bx)$ for brevity.
The first five terms are the known constraints, but rest of the terms
are not constraints. As shown, for pure Weyl gravity with $\Lambda=0$
and $\kappa^{-1}=0$ the result \eqref{pb:Q,HC} consists of
the first four terms which are all constraints. The extra terms that
are not constraints are a result of the fact that adding the
cosmological constant or the \EH action into Weyl gravity, breaks the
conformal symmetry.

Finally, we determine the Poisson bracket of the Hamiltonian constraint
with itself:
\begin{equation}
\begin{split}
\pb{\HC{\xi},\HC{\eta}}&=\int_{\Sigma_t}d^3x \left[
\pb{\HC{\xi},p^{ij}(\bx)} \pb{h_{ij}(\bx),\HC{\eta}} \right.\\
&\qquad +\left.\pb{\HC{\xi},\cP^{ij}(\bx)} \pb{K_{ij}(\bx),\HC{\eta}}
-(\xi\leftrightarrow\eta) \right] \\
&=\int_{\Sigma_t}d^3x \left[ F_{(1)}^i D_i\xi\eta
+F_{(2)}^{ij} D_{ij}\xi\eta +4\alpha\sqrt{h}D_i\left( \xi
C^{i(jk)}_{\phantom{k(ij)}\bn} \right) D_{jk}\eta \right.\\
&\qquad\quad -(\xi\leftrightarrow\eta) \bigr] \,,
\end{split}
\end{equation}
where we denote
\begin{equation}\label{F_(1)}
\begin{split}
F_{(1)}^i &=
2D^{j}\cP^{ik}K_{jk} +4D_j\cP^{jk}K^i_{\phantom ik}
-2D_j\cP^{ij}K -3D^i\cP^{jk}K_{jk} \\
&\quad -4C^i_{\phantom{i}jk\bn}\cP^{jk}
+4\alpha\sqrt{h}C^{ijk}_{\phantom{ijk}\bn} \sR_{jk}
\end{split}
\end{equation}
and
\begin{equation}\label{F_(2)}
F_{(2)}^{ij}=-2p^{ij} +4\cP^{(i|k}K^{|j)}_{\phantom{|j)}k} -2\cP^{ij}K
-3h^{ij}\cP^{kl}K_{kl} \,.
\end{equation}
It is noteworthy that this Poisson bracket is insensitive to the
presence of \EH or cosmological constant parts of the action. This means
that the result for the Poisson bracket between Hamiltonian constraints
does not contain the Hamiltonian constraint itself, but rather consists
of other constraints. In obtaining \eqref{F_(1)}, we used the fact that
the terms involving a product of three $K_{ij}$ turn out to be equal to
the covariant divergence of the characteristic polynomial
\eqref{P(A)=0.3d} of $K_{ij}$:
\begin{equation}\label{DK^3}
 C^{ijk}_{\phantom{ijk}\bn}K_{jl}K^l_{\phantom lk}
+K^i_{\phantom ij}C^{jkl}_{\phantom{jkl}\bn}K_{kl}
-C^{ijk}_{\phantom{ijk}\bn}K_{jk}K
=-D_jP(K)^{ij}=0 \,.
\end{equation}

Integrating by parts enables us to write the Poisson bracket between
Hamiltonian constraints as
\begin{equation}\label{pb:HC,HC}
\pb{\HC{\xi},\HC{\eta}}
=\int_{\Sigma_t}d^3x\left( \xi D_i\eta -D_i\xi\eta \right) G^i \,,
\end{equation}
where we denote
\begin{equation}
\begin{split}
G^i &=-F_{(1)}^i +D_jF_{(2)}^{ij}
+4\alpha\sqrt{h}C^{ijk}_{\phantom{ijk}\bn} \sR_{jk} \\
&=h^{ij}\left[ \cH_j +2\cP \left( D^kK_{jk} -D_jK \right) \right].
\end{split}
\end{equation}
The second- and third-order derivatives of the test functions cancel.
This is because $F_{(2)}^{ij}$ is symmetric and $C_{ijk\bn}$ inherits
the cyclic property of the Weyl tensor:
\begin{equation}\label{pCn-cyclic}
C_{ijk\bn}+C_{kij\bn}+C_{jki\bn}=0
\quad\Rightarrow\quad C_{(ijk)\bn}=0 \,.
\end{equation}
The Ricci identity was used in obtaining the coefficient of the
first-order derivatives of the test functions in \eqref{pb:HC,HC}, and
the Riemann tensor was written in terms of the Ricci tensor since the
three-dimensional Weyl tensor is zero, \eqref{Weyl-tensor-3d}.

Finally, we may write the Poisson bracket between Hamiltonian
constraints as a sum of the momentum and $\cP$ constraints,
\begin{equation}
\pb{\HC{\xi},\HC{\eta}}
=\MC{\xi\vec{D}\eta-\eta\vec{D}\xi}
+2\cP\bigl[ \left( \xi D_i\eta-\eta D_i\xi \right) h^{ij}
\bigl( D^kK_{jk}-D_jK \bigr) \bigr] \,,
\end{equation}
where the gradient vector $\vec{D}\xi$ is defined as
$(\vec{D}\xi)^i=h^{ij}D_j\xi$.

Substituting the test functions $\xi=\delta(\bx-\by)$ and $\eta=N$
gives the Poisson bracket between $\cH_0(\bx)$ and $\HC{N}$ as
a sum of the constraints $\cH_i$ and $\cP$,
\begin{equation}
\pb{\cH_0(\bx),\HC{N}}=\left( 2D^iN+ND^i \right)
\left[ \cH_i +2\cP \left( D^jK_{ij}-D_iK \right) \right](\bx) \,.
\end{equation}

\subsection{Curvature-squared gravity\texorpdfstring{:
$\alpha\neq0$, $\beta\neq0$, $\gamma=0$}{}}\label{appendix4.2}
We consider the case when both the Weyl tensor squared and scalar
curvature squared terms are included in the Lagrangian \eqref{S_C}.
The \EH action and the cosmological constant may be included or
excluded, since they do not alter the kinetic part of the Hamiltonian
constraint. The Hamiltonian constraint is given in \eqref{H_0.R2}. When
the cosmological constant and/or the \EH action are not present in the
action, one simply sets $\Lambda=0$ and/or $\kappa^{-1}=0$ in the
following results.

\subsubsection{Poisson brackets between the canonical variables and the
Hamiltonian constraint}
The Poisson bracket between $K_{ij}$ and the Hamiltonian constraint
$\HC{\xi}$ reads as
\begin{multline}
\pb{K_{ij}(\bx),\HC{\xi}}= \xi(\bx)\left[
-\frac{\cG_{ijkl}\cP^{kl}}{\alpha\sqrt{h}} +\sR_{ij} +K_{ij}K
-\frac{1}{2}h_{ij} \left( \sR -K_{ij}K^{ij} +K^2 \right) \right](\bx) \\
+D_{ij} \xi(\bx)\,.
\end{multline}

The Poisson bracket between the momentum $\cP^{ij}$ and the Hamiltonian
constraint $\HC{\xi}$ is obtained as
\begin{equation}
\begin{split}
\pb{\HC{\xi},\cP^{ij}(\bx)} &= \xi(\bx) \left[ 2p^{ij}
+\cP^{ij}K +h^{ij}\cP^{kl}K_{kl} +\cP\left( K^{ij}-h^{ij}K \right)
 \right.\\
&\quad -\left. \frac{\sqrt{h}}{\kappa}\left( K^{ij}-h^{ij}K \right)
 \right](\bx)
+4\alpha\sqrt{h}D_k\left( \xi C^{k(ij)}_{\phantom{k(ij)}\bn}
\right)(\bx) \,.
\end{split}
\end{equation}

The Poisson bracket between the momentum $p^{ij}$ and the Hamiltonian
constraint $\HC{\xi}$ is obtained via a similar calculation as in the
case with $\beta=0$. We obtain it as
\begin{equation}\label{pb:HC,p^ij.R2}
\pb{\HC{\xi},p^{ij}(\bx)}= E_{(0)}^{ij} \xi(\bx) +
E_{(1)}^{ijk}D_k\xi(\bx) +E_{(2)}^{ijkl}D_{kl}\xi(\bx) \,,
\end{equation}
where we have defined the three coefficient tensor densities
$E_{(I)}^{i_1\cdots i_{2+I}}$ ($I=0,1,2$) as
\begin{equation}
\begin{split}
E_{(0)}^{ij}&=-\frac{1}{\alpha\sqrt{h}}\left(
\cP^i_{\phantom{i}k}\cP^{jk}
-\frac{\alpha+3\beta}{9\beta}\cP^{ij}\cP
 -\frac{1}{4}h^{ij}\cP^{kl}\cG_{klmn}\cP^{mn} \right) \\
&\quad +D_kD^{(i}\cP^{j)k}
-\frac{1}{2}h^{ij}D_kD_l\cP^{kl}
-\frac{1}{2}D^kD_k\cP^{ij}
 -\frac{1}{2}D^{ij}\cP +\frac{1}{2}h^{ij}D^kD_k\cP \\
&\quad  -\cP^{kl}K_{kl}K^{ij}
-\frac{1}{2}\cP^{ij} \left( \sR -K_{kl}K^{kl} +K^2 \right) \\
&\quad +\frac{1}{2}\cP\sR^{ij} -\cP K^i_{\phantom ik}K^{jk}
+\cP K^{ij}K  -\frac{1}{2}\sqrt{h}h^{ij}\Lambda \\
&\quad +\frac{\sqrt{h}}{2\kappa}\left[ \sR^{ij}
+2K^i_{\phantom ik}K^{jk}-2K^{ij}K
-\frac{1}{2}h^{ij}\left( \sR+K_{ij}K^{ij}-K^2 \right) \right] \\
&\quad +\alpha\sqrt{h} \left[
+2C^i_{\phantom{i}kl\bn}C^{jkl}_{\phantom{jkl}\bn}
+C_{kl\phantom{i}\bn}^{\phantom{kl}i}
C^{klj}_{\phantom{klj}\bn}
-\frac{1}{2}h^{ij}C_{klm\bn}C^{klm}_{\phantom{klm}\bn} \right.\\
&\quad -2D_kK_l^{\phantom l(i}C^{j)lk}_{\phantom{j)lk}\bn}
-2D_kK_l^{\phantom l(i|}C^{kl|j)}_{\phantom{kl|j)}\bn}
-2K_l^{\phantom l(i}D_k C^{j)lk}_{\phantom{j)lk}\bn}\\
&\quad -\left.2K_l^{\phantom l(i|}D_k C^{kl|j)}_{\phantom{kl|j)}\bn}
-2K_{kl}D^k C^{l(ij)}_{\phantom{l(ij)}\bn}
-2C^{k(ij)}_{\phantom{k(ij)}\bn}\left(2D_lK^l_{\phantom lk}
-D_kK\right) \right],
\end{split}
\end{equation}
\begin{equation}
\begin{split}
E_{(1)}^{ijk}&=D^{(i}\cP^{j)k} +2h^{k(i}D_l\cP^{j)l}
-h^{ij}D_l\cP^{kl} -\frac{3}{2}D^k\cP^{ij}
-D^{(i}\cP h^{j)k} +h^{ij}D^k\cP \\
&\quad -2\alpha\sqrt{h}\left( K_l^{\phantom l(i}
C^{j)lk}_{\phantom{j)lk}\bn}
+K_l^{\phantom l(i|}C^{kl|j)}_{\phantom{kl|j)}\bn}
+K^k_{\phantom kl}C^{l(ij)}_{\phantom{l(ij)}\bn} \right)
\end{split}
\end{equation}
and
\begin{equation}
\begin{split}
E_{(2)}^{ijkl}&=h^{i(k}\cP^{l)j} +h^{j(k}\cP^{l)i}
-\frac{1}{2}h^{ij}\cP^{kl} -h^{kl}\cP^{ij}
+\frac{1}{2}\cP \left( h^{ij}h^{kl} -h^{i(k}h^{l)j} \right) \\
&\quad +\frac{\sqrt{h}}{2\kappa}\left( h^{ij}h^{kl}
-h^{i(k}h^{l)j} \right)  \,.
\end{split}
\end{equation}

\subsubsection{Poisson bracket between the Hamiltonian constraints}
We determine the Poisson bracket between Hamiltonian constraints:
\begin{equation}
\begin{split}
\pb{\HC{\xi},\HC{\eta}}
&=\int_{\Sigma_t}d^3x \left[ F_{(1)}^i D_i\xi\eta
+F_{(2)}^{ij} D_{ij}\xi\eta +4\alpha\sqrt{h}D_i\left( \xi
C^{i(jk)}_{\phantom{k(ij)}\bn} \right) D_{jk}\eta \right.\\
&\qquad\quad -(\xi\leftrightarrow\eta) \bigr] \,, \\
\end{split}
\end{equation}
where we denote
\begin{equation}
\begin{split}
F_{(1)}^i &= 2D^{j}\cP^{ik}K_{jk} +4D_j\cP^{jk}K^i_{\phantom ik}
-2D_j\cP^{ij}K -3D^i\cP^{jk}K_{jk} \\
&\quad -4C^i_{\phantom{i}jk\bn}\cP^{jk}
-2D^j\cP K^i_{\phantom ij} +2D^i\cP K
+4\alpha\sqrt{h}C^{ijk}_{\phantom{ijk}\bn}\sR_{jk}
\end{split}
\end{equation}
and
\begin{equation}
F_{(2)}^{ij}=-2p^{ij} +4\cP^{(i|k}K^{|j)}_{\phantom{|j)}k} -2\cP^{ij}K
-3h^{ij}\cP^{kl}K_{kl} +2\cP\left( h^{ij}K -K^{ij} \right).
\end{equation}
We again integrate by parts to obtain \eqref{pb:HC,HC}, but now with
$G^i=h^{ij}\cH_j$. Thus the result is given solely by the momentum
constraint:
\begin{equation}
\begin{split}
\pb{\HC{\xi},\HC{\eta}}&=\int_{\Sigma_t}d^3x\left( \xi D_i\eta-\eta
D_i\xi \right) h^{ij} \cH_j \\
&=\MC{\xi\vec{D}\eta-\eta\vec{D}\xi} \,.
\end{split}
\end{equation}
This result has the same form as in the Hamiltonian structure of general
relativity. Finally, we obtain
\begin{equation}
\pb{\cH_0(\bx),\HC{N}}=2D^iN\cH_i +ND^i\cH_i \,.
\end{equation}

\end{document}